\documentclass[a4paper,twocolumn,showpacs,superscriptaddress,pra]{revtex4}

\usepackage[english]{babel}
\usepackage[utf8]{inputenc}
\usepackage{graphicx}
\usepackage{latexsym}
\usepackage{amsfonts} 
\usepackage{amsmath}

\def\ket|#1>{| #1 \rangle}
\def\bra<#1|{\langle #1 |}
\def\<{\langle}
\def\>{\rangle}
\def\{{\lbrace}
\def\}{\rbrace}
\def\({\left(}
\def\){\right)}
\def\[{\left[}
\def\]{\right]}

\begin{document}

\title{Entanglement in low-energy states of the random-hopping model}%

\author{Giovanni Ram\'{\i}rez}%
\email{giovanni.ramirez@uam.es}%
\affiliation{Instituto de Física Teórica, UAM-CSIC, Madrid, Spain}%

\author{Javier Rodr\'{\i}guez-Laguna}%
\affiliation{Mathematics Dept., Universidad Carlos III de
  Madrid, Spain}%

\author{Germ\'an Sierra}
\affiliation{Instituto de Física Teórica, UAM-CSIC, Madrid, Spain}%

\date{Mar. 7, 2014}

\begin{abstract}
  We study the low-energy states of the 1D random-hopping model in the strong
  disordered regime. The entanglement structure is shown to depend solely on
  the probability distribution for the length of the effective bonds $P(l_b)$,
  whose scaling and finite-size behavior are established using
  renormalization-group arguments and a simple model based on random
  permutations. Parity oscillations are absent in the von Neumann entropy with
  periodic boundary conditions, but appear in the higher moments of the
  distribution, such as the variance. The particle-hole excited states leave
  the bond-structure and the entanglement untouched. Nonetheless, particle
  addition or removal deletes bonds and leads to an effective saturation of
  entanglement at an effective block size given by the expected value for the
  longest bond.
\end{abstract}


\maketitle

\section{Introduction}

The interplay between disorder and entanglement in low-dimensional systems has
proved to be a rich source of problems and surprises. Anderson's theorem
\cite{Anderson.58} states that, in one-dimensional systems with uncorrelated
disorder in the local potential, all single-body states will localize and,
thus, real space blocks within the ground state present nearly no
entanglement. Entanglement entropy is, nonetheless, a good indicator of the
localization-delocalization quantum phase transition in higher dimensions
\cite{Chakravarti.PRB.08}. On the other hand, off-diagonal disorder, as it
appears in the random variants of the XX or Ising models, leads in certain
cases to long-range correlations and logarithmic violations of the area law.

Let us focus on a relevant special case in which the clean system is in a 1D
critical state, described by a certain conformal field theory (CFT) with
central charge $c$. The von Neumann entropy of a block of $\ell$ contiguous
sites in a system of size $L$ with periodic boundary conditions follows the
law \cite{Holzhey.94, Vidal_etal.03, Calabrese.JPA.09}
\begin{equation}
  \label{S.cft}
  S(\ell) \approx {c\over 3} \log{(\ell)}.
\end{equation}

There is ample evidence that the inclusion of strong off-diagonal disorder
between nearest neighbors gives raise to a disorder-averaged von Neumann
entropy similar to eq. (\ref{S.cft}), but with a different {\em effective}
value for the central charge \cite{Refael_Moore.JPA.09}:
\begin{equation}
  \label{S.disorder}
  \<S(\ell)\> \approx {c \log{(s)}\over 3} \log{(\ell)} + c',
\end{equation}
where $\log{(s)}$ is the von Neumann entropy of the ground state of a system
with two sites. Indeed, the striking similarities between the clean and the
strongly disordered systems are even deeper than expression (\ref{S.disorder})
suggests, since they also appear in the averages for the correlation functions
and the finite-size effects in entanglement \cite{Fagotti.PRB.11}. Those
similarities will be the main focus of this work.

In this work we will consider the random-XX model or, in other terms, the
fermionic random-hopping Hamiltonian in 1D:
\begin{equation} 
  \label{random.hop}
  {\cal H} = -\sum_i J_i c^\dagger_i c_{i+1} + \hbox{h.c.}.
\end{equation}

In the clean case, the system is critical, and the central charge of the
associated CFT is $c=1$. The $\log(s)$ factor is found by considering what is
the entanglement entropy between sites in a $L=2$ system, i.e.:
$\log(s)=\log(2)$. Thus, we can fill in the values $c=1$ and $s=2$ in
expression (\ref{S.disorder}).

Whenever the $J_i$ are different, expression (\ref{random.hop}) is called the
{\em inhomogeneous hopping} model, which is exactly solvable: diagonalizing
the hopping matrix one can study the single particle energy levels
$\epsilon_k$ and the single particle modes $v_{k,i}$, where $k$ denotes the
eigenvalue number and $i$ the actual site. When the different {\em hoppings}
$J_i$ vary slowly with position, they can be regarded as a modulation on the
speed of sound. Indeed, a careful choice for the $J_i$ can be used to model
quantum matter on a curved space-time background \cite{Boada.NJP.11}. The
modes have some generic mathematical properties, such as particle-hole
symmetry: a canonical transformation $c^\dagger_i \to (-1)^i c^\dagger_i$
transforms ${\cal H}\to -{\cal H}$. Thus, if $\{v_{k,i}\}_{i=1}^L$ is a mode
with energy $\epsilon_k$, then $\{(-1)^iv_{k,i}\}_{i=1}^L$ is also a mode with
energy $-\epsilon_k$. In absence of zero modes, the ground state can be proved
to take place at half filling and is spatially homogeneous.

Let us consider the $\{J_i\}$ to be independent random variables extracted
from a probability distribution $p_\delta(J)$ pertaining to the following
family
\begin{equation}
  \label{prob.delta}
  p_\delta(J)\equiv {1\over\delta} J^{-1+{1\over\delta}}
\end{equation}
for $0<J<1$ and $\delta>0$; $\delta$ characterizes the randomness,
i.e. $\delta=1$ for the uniform distribution. We will focus on the
$\delta\to\infty$ limit, the so-called strong disorder regime
\cite{Refael_Moore.JPA.09}, in which the sampled $J$ span many orders of
magnitude in the interval $(0,1)$. In this regime, the renormalization group
(RG) derived by Dasgupta and Ma \cite{Dasgupta_Ma.80} gives an accurate
description of the ground state. It proceeds through decimation: at each RG
step, we select the highest hopping and put a fermion resonating between the
two sites, i.e.: a {\em bond}. In the spin-chain view, we would speak of a
singlet between both spins. The two sites are then removed from the chain, and
their neighbors are linked by a new effective hopping term. The procedure is
repeated and, when it is finished, the ground state can be written as a {\em
  random bond} (or random singlet) structure. Under successive applications of
the Dasgupta-Ma RG approach, the probability distribution for the remaining
hoppings, $p_\delta(J)$ flows by increasing the value of $\delta$, and
$\delta\to\infty$ is the (unattainable) infinite-randomness fixed point (IRFP)
\cite{Fisher.PRB.95,Fisher.PRB.98}. That is the reason for our choice to focus
on the strong disorder limit.

The aim of this work is to illuminate the surprising relation between
entanglement in critical states, as described by CFT, and average entanglement
entropies in strongly disordered systems. In section \ref{sec:computing} we
will discuss the techniques which will be employed to obtain expectation
values of entanglement measures, using both exact diagonalization and the
RG. In section \ref{sec:gs} we will discuss the finite-size scaling of the
average von Neumann and R\'enyi block entropies within the ground state of our
model. The statistics of the bond lengths is discussed in detail in section
\ref{sec:bondpic}. By successive distillation of the basic physics we will
reach a simple model which yields accurate predictions for average entropies,
based solely on the analysis of random permutations in section
\ref{sec:perm}. The study of the excited states, performed in section
\ref{sec:exc}, benefits from the different approaches discussed
before. Section \ref{sec:conclusions} is devoted to the exposition of
conclusions and further work.

\section{Computing entanglement entropies}
\label{sec:computing}

Our aim is to study the statistical properties of the entanglement of the
ground state of Hamiltonian (\ref{random.hop}) on a 1D system with size $L$
and periodic boundary conditions (PBC), when the couplings $\{J_i\}$ are
chosen as independent random variables from the probability distribution
(\ref{prob.delta}), specially in the strong disorder limit
$\delta\to\infty$. The physics of the low-energy eigenstates of Hamiltonian
(\ref{random.hop}) can be analyzed with two methods:

{\em 1.- Exact Diagonalization (ED)}. In order to obtain the ground state of
(\ref{random.hop}) it suffices to diagonalize the hopping matrix $T_{ij}$,
with $i,j\in \{1,\cdots,L\}$, whose non-zero elements are $T_{i,i+1}
=T_{i+1,i} =-J_i$. Its eigenstates, $T v_k =\epsilon_k v_k$, constitute the
modes, which allow us to compute the correlation matrix:
\begin{equation}
  \label{correlator}
  C_{ij}\equiv \<c^\dagger_i c_j\> = \sum_{k=1}^{N_e} \bar v_{k,i} v_{k,j},
\end{equation}
for the ground state of the system: a Fermi state of $N_e$ fermions. Given any
block $B$ of $\ell$ sites in the system, the eigenvalues of the reduced
density matrix, i.e.: its entanglement spectrum, can be obtained using a
reverse form of Wick's theorem \cite{Peschel.JPA.03}. Let $\nu_k$ be the
eigenvalues of the correlation matrix when restricted to that block, then the
reduced density matrix $\rho^B$ of the block is a product of density matrices
of single-site blocks $\bigotimes_k^\ell\rho_k$ and the $\alpha$-th order
R\'enyi entropy can be obtained as a sum of the entropies of $\rho_k = \nu_k
d_k^\dagger d_k + (1-\nu_k) d_k d_k^\dagger$ where $d$ and $d^\dagger$ are
other fermionic operators
\begin{equation}
  \label{renyi}
  S_\alpha(B)={1\over 1-\alpha} \sum_k \[ \nu_k^\alpha +
  (1-\nu_k)^\alpha \]
\end{equation}
and the von Neumann entropy for the block corresponds to the limit $\alpha \to
1$,
\begin{equation}
  \label{von_neumann}
  S_1(B)= -\sum_k \[\nu_k \log{(\nu_k)} + (1-\nu_k)
  \log{(1-\nu_k)} \].
\end{equation}

The application of this technique in our case presents a serious problem: the
diagonalization of the hopping matrix is a highly ill-conditioned problem,
since their eigenvalues differ by many orders of magnitude. Thus, it is a
numerical challenge to obtain the actual block entropy, given a realization of
the $\{J_i\}$. In practice, it is unfeasible to study systems with either $L$
or $\delta$ too large.

{\em 2.- Dasgupta-Ma Renormalization Group (RG)}. Within the strong disorder
regime, we can rely on the renormalization group (RG) scheme devised by
Dasgupta and Ma \cite{Dasgupta_Ma.80} to find out the bond structure which
describes the ground state of the system. This is a decimation procedure in
which one chooses the strongest link, $\max \{J_i\}$ and establishes a
single-particle state as a bond on top of it. Then, the two neighboring sites
are joined by a renormalized (effective) link, whose strength can be found
using second-order perturbation theory:
\begin{equation}
  \label{renormalized.coupling}
  J^{(R)}_k = {J_{k-1}J_{k+1}\over J_{k}}.
\end{equation}

\begin{figure}
  \centering
  \resizebox{6.0cm}{!}{\includegraphics{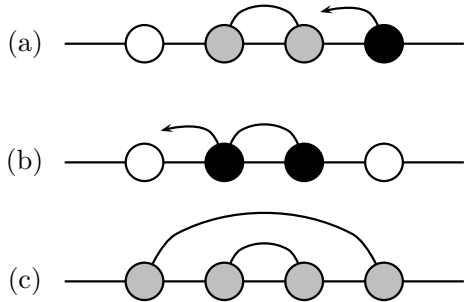}}
  \caption{Illustration of the physical picture which induces long-distance
    bonds. (a) A bond has been established on the central link, which is very
    strong. A particle at the right extreme attempts to jump in. (b)
    Sometimes, the particle succeeds, and the central bond becomes doubly
    occupied. The left particle must jump out. (c) We can view the full
    procedure as a tunneling event through the occupied bond, with a much
    lower associated probability amplitude.}
  \label{fig.illust}
\end{figure}

The strongest link and its two neighbours are replaced by this (weaker)
renormalized link, then we proceed to pick the second strongest link, and
iterate the process until all the links have been renormalized (assuming even
$L$). At some moment, the strongest link will be one of the renormalized links
in previous iterations. Thus, a long-distance bond will be established between
two sites which were not nearest neighbors. Why do such long-distance bonds
exist? The physical picture is illustrated in figure \ref{fig.illust}. Let us
consider the particle at the rightmost site. It has a certain probability of
hopping to its left, whenever the inner bond particle is also at its left
site. At this moment, the inner bond becomes doubly occupied. The original
particle inside the inner bond is not allowed to hop rightwards, but it may
hop leftwards. As particles are indistinguishable, the total procedure can be
described as a tunneling of one particle through an established bond. The
associated probability amplitude of this event is much lower than the
probability amplitude of hopping in the inner bond, thus accounting for the
large differences in energy between them. This procedure, which is akin to the
Anderson mechanism describing the interaction between a magnetic impurity and
the spin of a conduction electron, can be assigned an effective hopping
amplitude using second-order perturbation theory, thus obtaining expression
(\ref{renormalized.coupling}). Similarly, one can think of the second-order
procedure which allows to find an effective Heisenberg Hamiltonian with
$J\approx t^2/U$ from a Hubbard system in the limit $U\gg t$.

\begin{figure}
  \centering
  \resizebox{4.0cm}{!}{\includegraphics{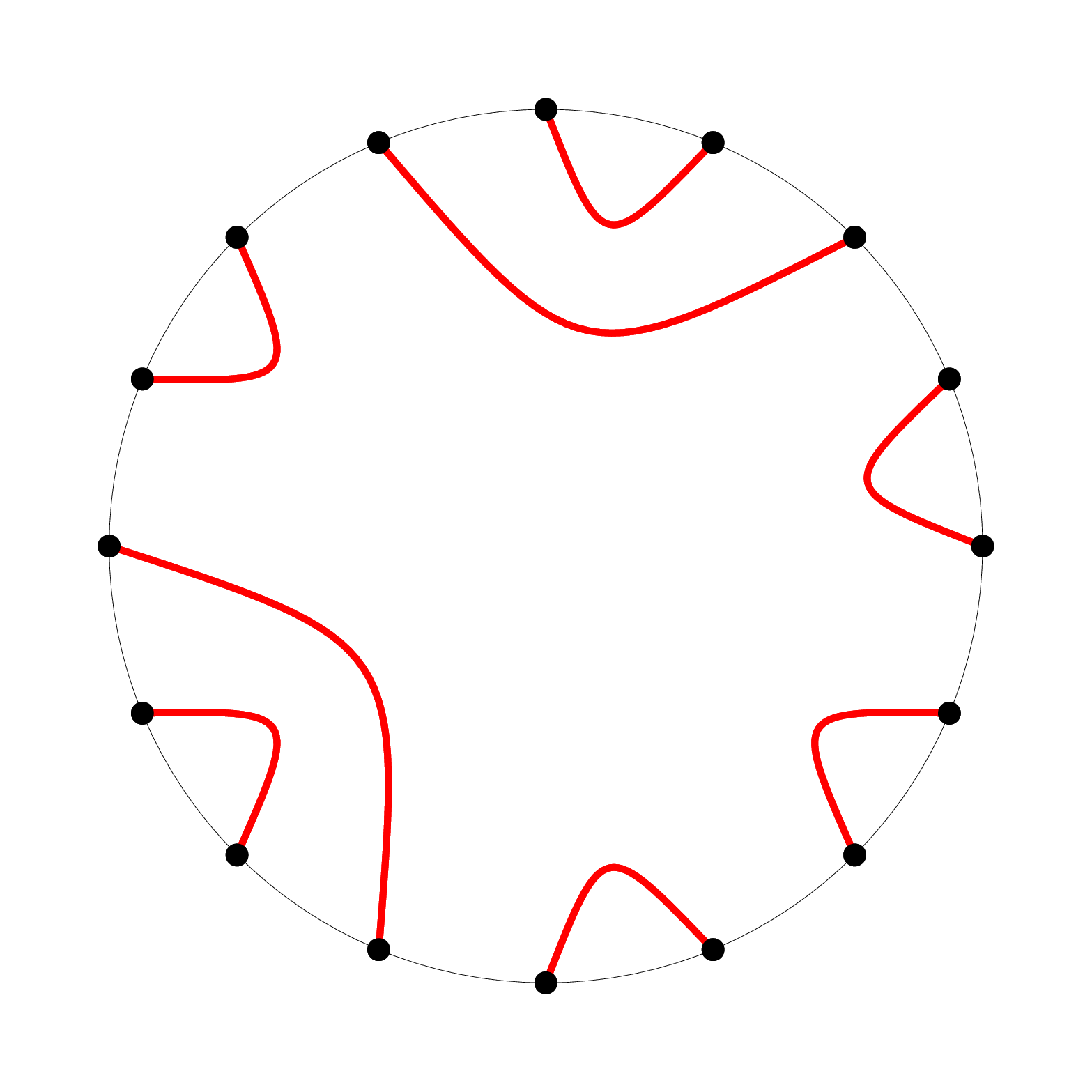}}%
  \resizebox{4.0cm}{!}{\includegraphics{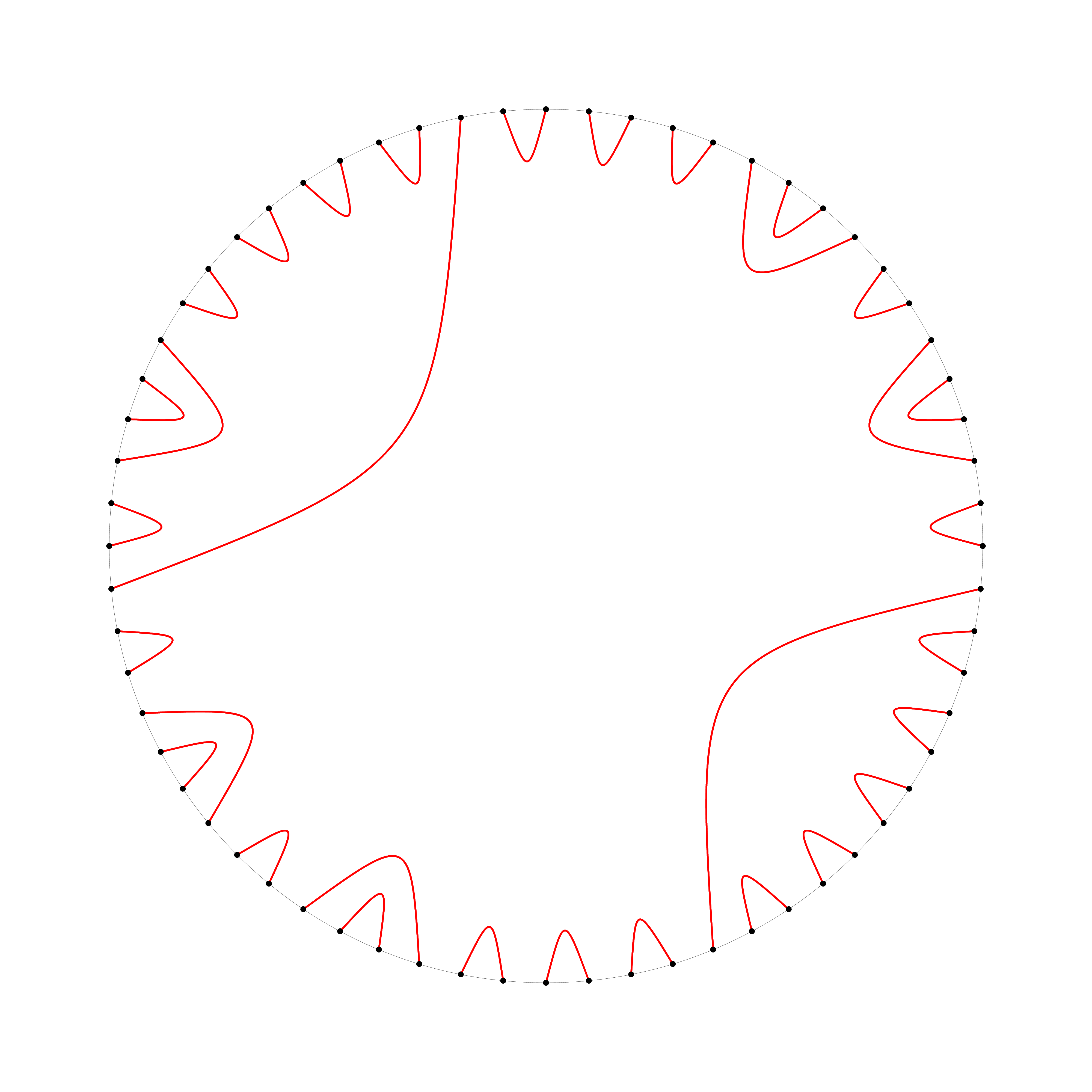}}
  \caption{Example of bond structures, with $16$ (left) and $64$ sites (right)
    with periodic boundary conditions and $\delta=8$.}
  \label{fig.bonds}
\end{figure}

When the decimation procedure is finished, we obtain a bond-structure, as
those illustrated in figure (\ref{fig.bonds}), with many bonds of length one,
but still with a large fraction covering larger distances. Notice that the
random-bond state factorizes into pairs, i.e.: there is a pairing of the sites
$\{(i_1,j_1), (i_2,j_2), \cdots, (i_{L/2},j_{L/2})\}$, such that the ground
state for the system factorizes into the product of a singlet state for every
pair. In other words, each pair $(i_k,j_k)$ is disentangled from the rest of
the system. The entanglement entropy of any block $B$ can be found by simply
counting the number of bonds which connect $B$ to the rest of the system, and
multiplying by $\log(2)$, which is the entropy associated to a single
bond. Thus, the RG opens the possibility of a purely combinatorial solution to
this problem, which will be discussed in section \ref{sec:perm}. Notice that,
within the bond picture, all R\'enyi entropies are equal.

Remarkably, the Dasgupta-Ma RG has recently received an interpretation within
the the tensor-networks and holography language \cite{Goldsborough.14}.

\section{Average Entanglement Entropies in the Ground State}
\label{sec:gs}

\begin{figure}[h!]
  \centering
  \resizebox{8.0cm}{!}{\includegraphics{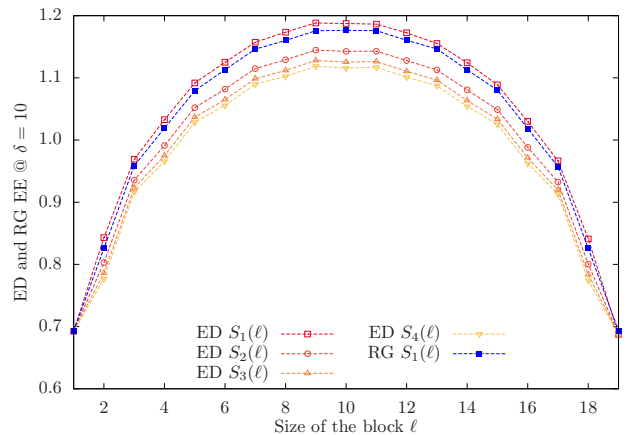}}
  \caption{Average von Neumann and R\'enyi block entropies for a $L=20$ system
    with $\delta=10$, comparing exact diagonalization and RG results for
    $10^6$ realizations. Notice that the RG gives the same curve for all
    R\'enyi orders, which is closest to the von Neumann entropy obtained by
    exact diagonalization.}
  \label{fig.diag}
\end{figure}

Let $S_\alpha(\ell)$ be the disorder-averaged R\'enyi entropy of a block of
size $\ell$ of order $\alpha$. When $\alpha=1$, i.e., the von Neumann entropy,
we will sometimes drop the index. Figure \ref{fig.diag} compares the averaged
R\'enyi entropies with both methods, exact diagonalization and RG, for a chain
with PBC, $L=20$ and $\delta=10$. The RG approach yields the same curve for
all orders $S_\alpha$, while they differ for exact diagonalization. Notice
that the RG entropy is closest to the $S_1$ exact diagonalization entropy.
 
\begin{figure}[h!]
  \centering
  \resizebox{4.0cm}{!}{\includegraphics{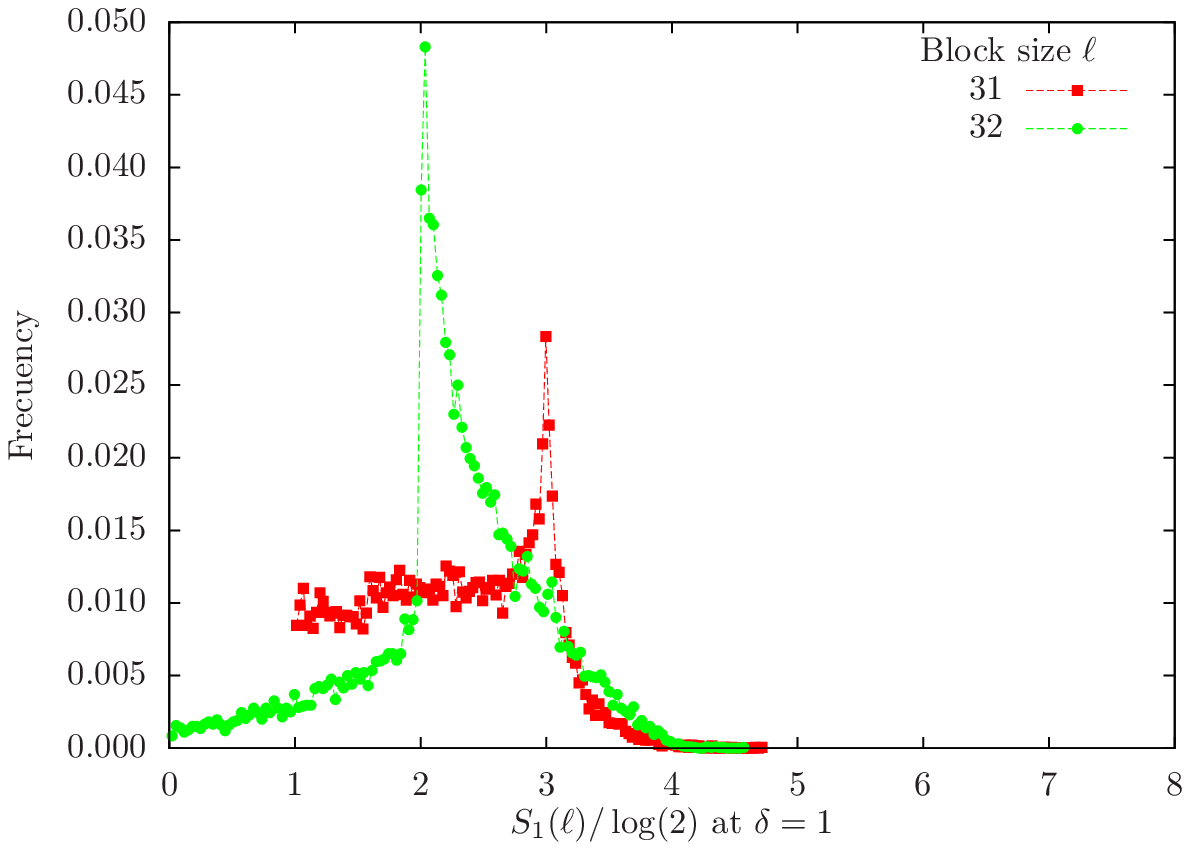}}%
  \resizebox{4.0cm}{!}{\includegraphics{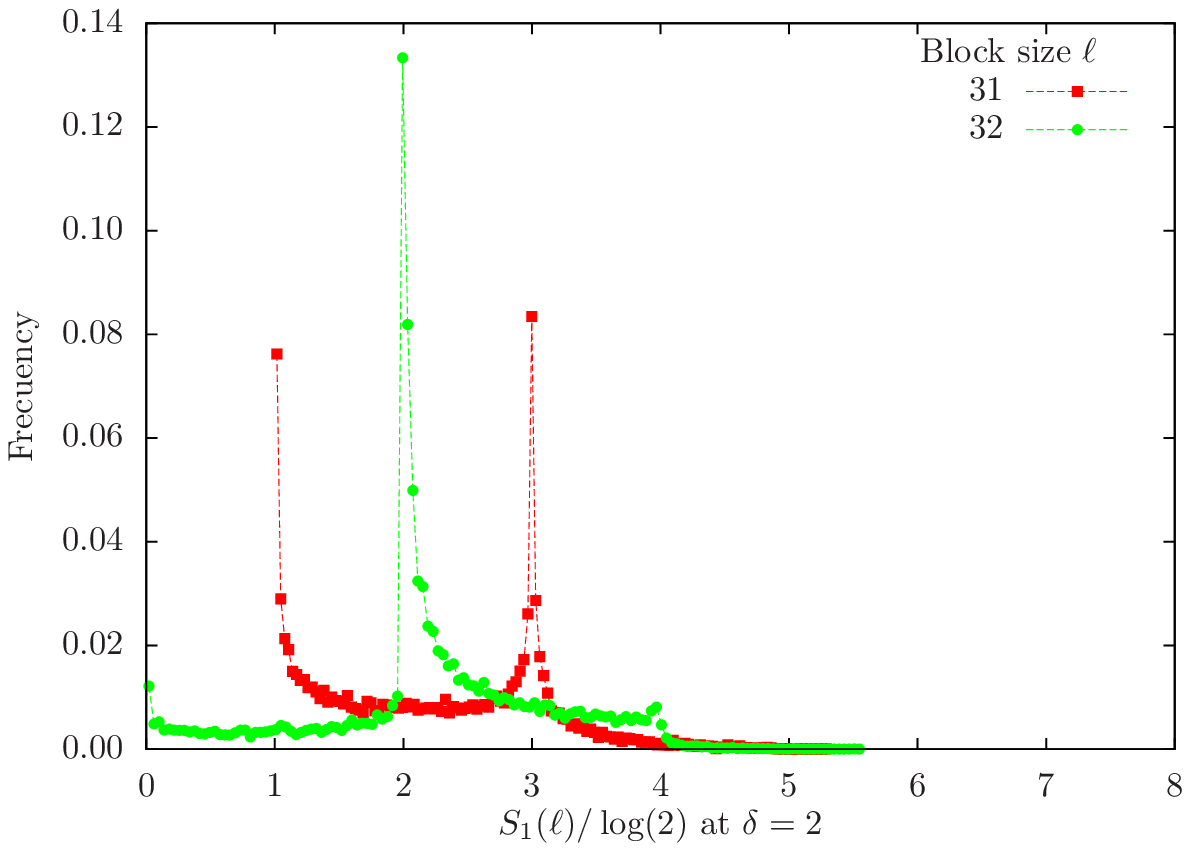}}
  \resizebox{4.0cm}{!}{\includegraphics{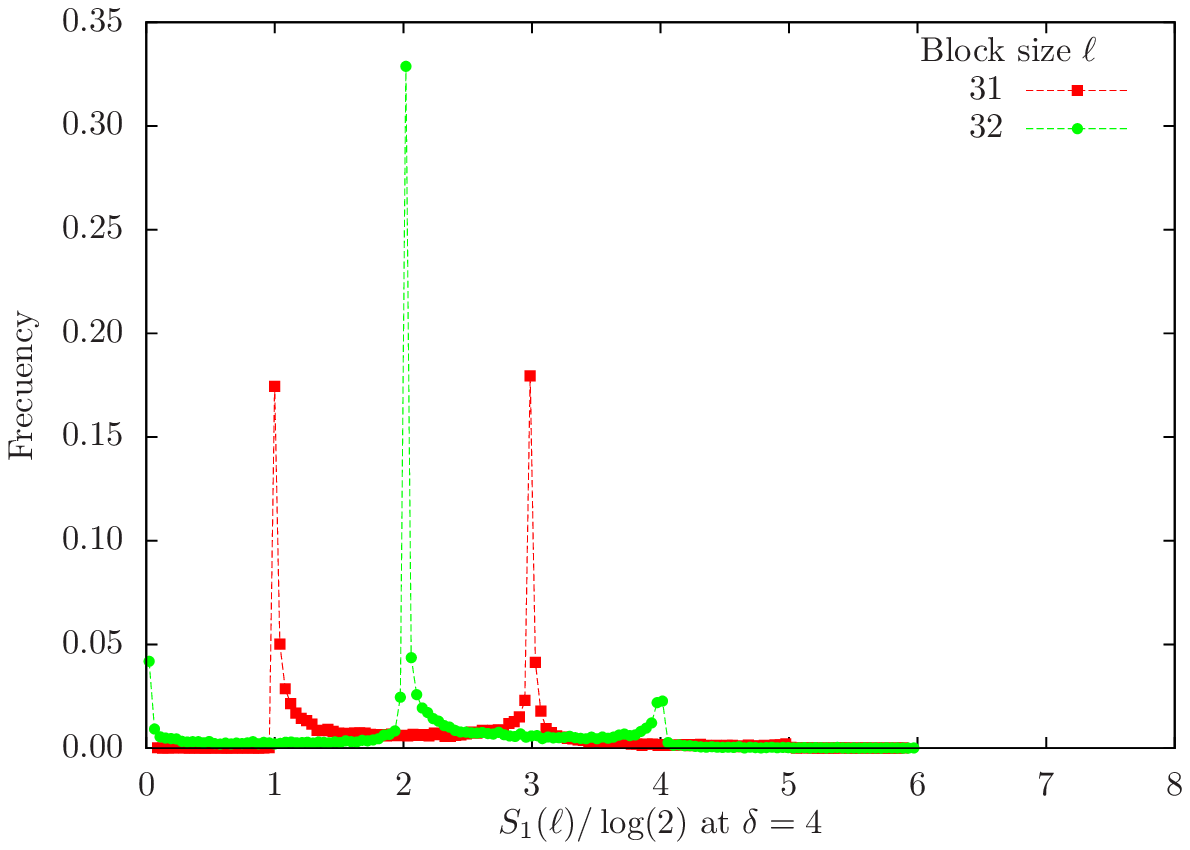}}%
  \resizebox{4.0cm}{!}{\includegraphics{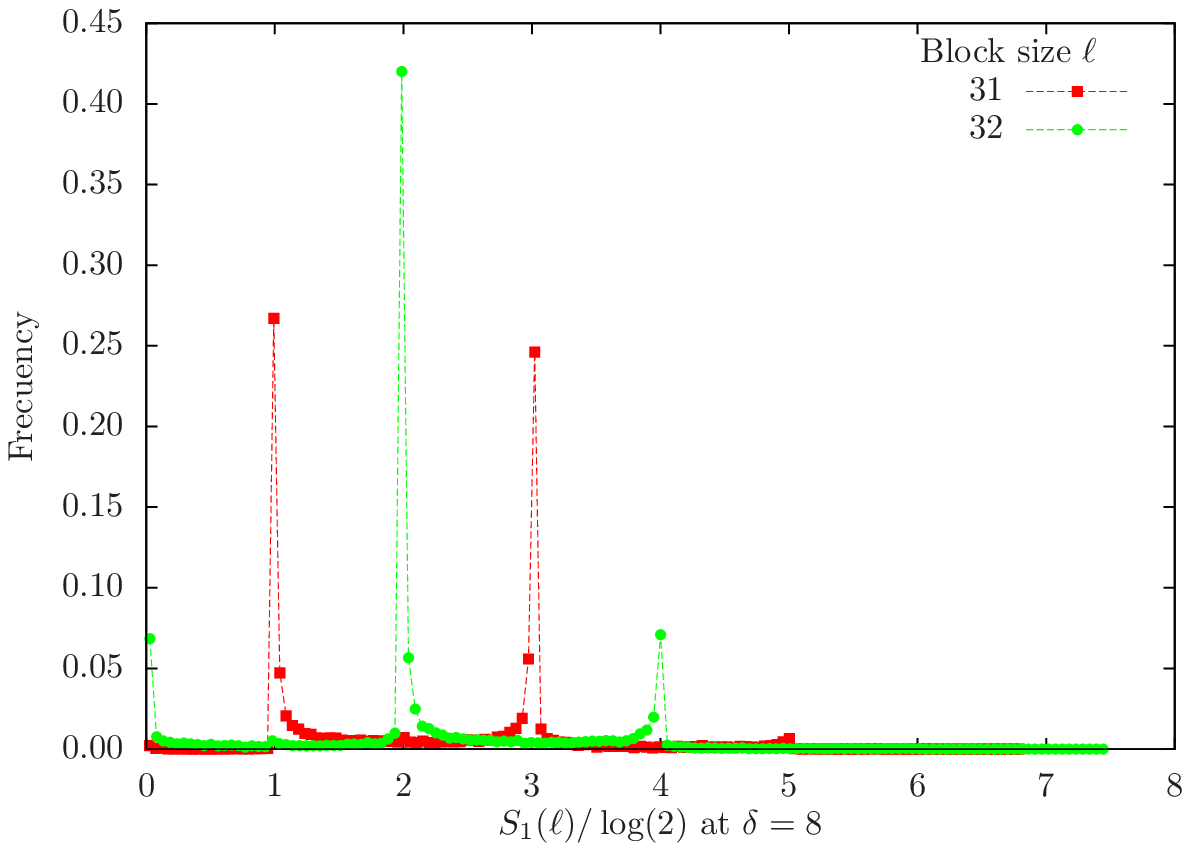}}
  \caption{Histogram for the von Neumann entropy for odd-size blocks (red) and
    even-size blocks (green) for different values of $\delta$ and $L=64$ for
    $2\cdot 10^4$ samples.}
  \label{fig.fluct}
\end{figure}

The difference between the RG predictions and the exact diagonalization
results can be ascribed to inaccuracies in the bond-structure picture. Figure
(\ref{fig.fluct}) shows a histogram of the values of the von Neumann entropy
at half-chain for different disorder realizations. Notice that, as $\delta$
increases, the behavior becomes closer to the bond-structure picture, which
predicts a set of delta peaks at integer multiples of $\log(2)$
\cite{Laflorencie.PRB.05}.

\begin{figure}[h!]
  \centering
  \resizebox{8.0cm}{!}{\includegraphics{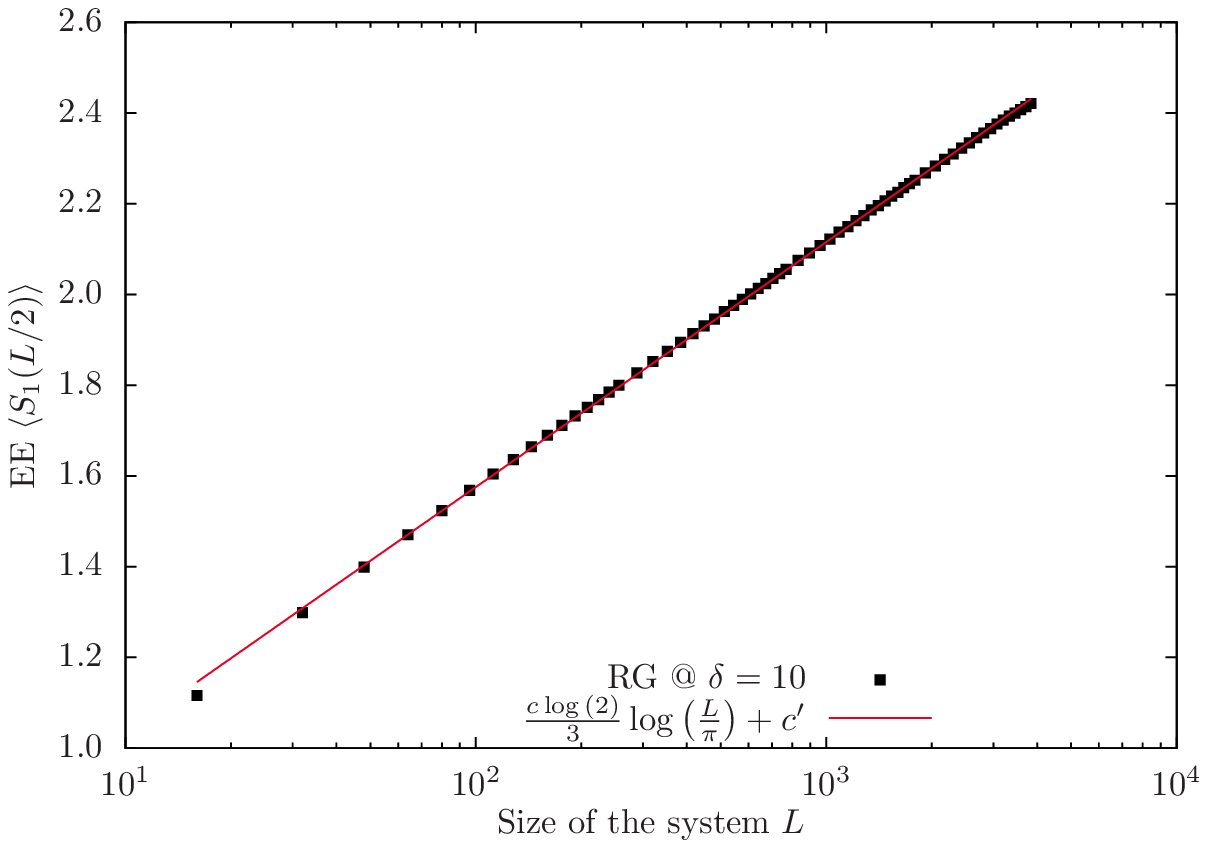}}
  \resizebox{8.0cm}{!}{\includegraphics{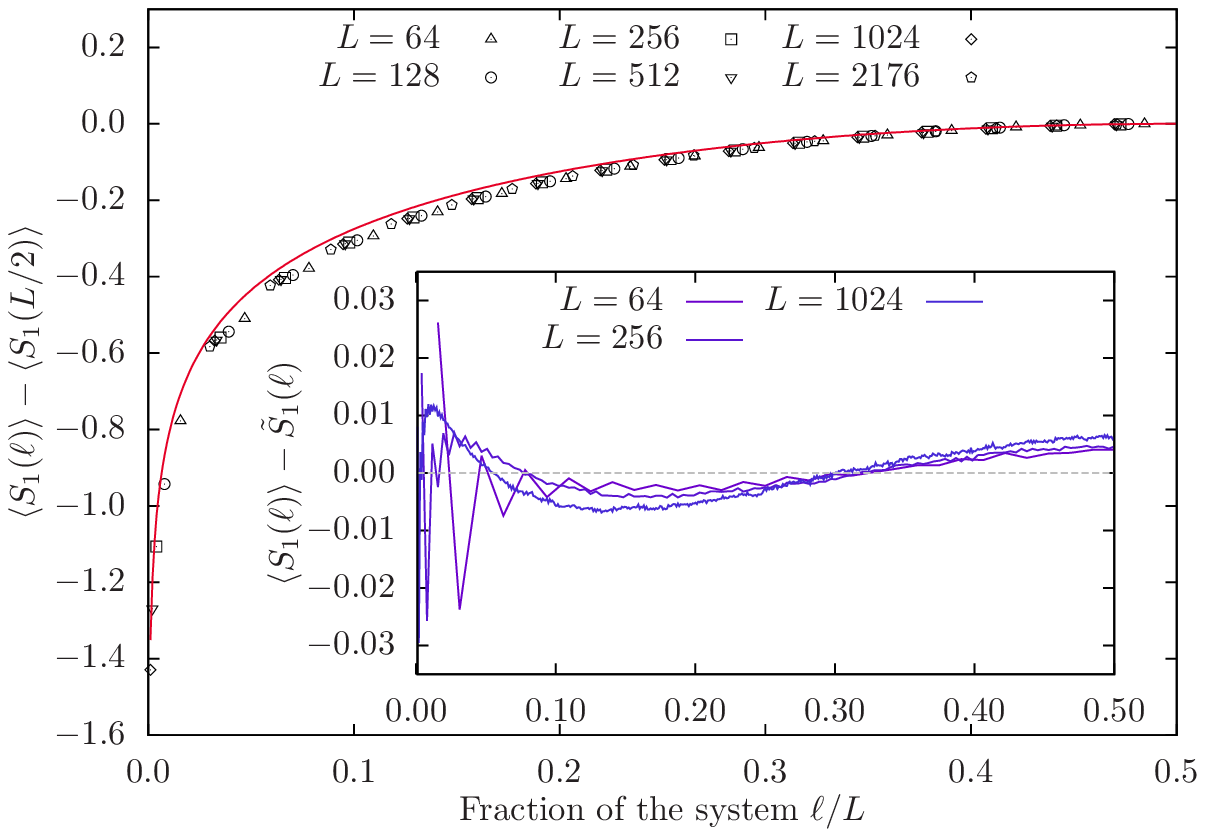}}
  \caption{Top: average von Neumann entropy at half-chain $\langle
    S_1(L/2)\rangle$ showing the characteristic logarithmic scaling with
    prefactor close to $\log(2)/3$. Bottom: vertically shifted data $\langle
    S_1(\ell)\rangle -\langle S_1(L/2)\rangle$ for different system sizes $L$
    collapse into the continuous line, fitting $\tilde S(\ell)$. Inset:
    residual error when the fitted expression is compared with the data, note
    the presence of higher harmonics.}
  \label{fig.Sgs}
\end{figure}

The average half-chain von Neumann entropy $\<S_1(L/2)\>$ is specially useful
to determine the global behavior disregarding finite-size effects. We have run
five million realizations of the disorder with $\delta=10$ and $L$ in the
range from $64$ to $2176$, and obtained the average half-chain entropy as a
function of $L$ using the Dasgupta-Ma RG, as shown in the top panel of
fig. \ref{fig.Sgs}. The fit to a form like (\ref{S.disorder}) is very accurate
\cite{Laflorencie.PRB.05,Refael.PRL.04}: $\<S_1(L/2)\>$ grows logarithmically
with a factor $c\log(2)/3$, and the fit for $c\approx 1.015$, i.e.: very close
to $1$. The additive constant is $c'\approx 0.7639$.

The average entropy for blocks of different sizes is shown in figure
(\ref{fig.Sgs}, bottom). We depict $\<S_1(\ell)\>-\<S_1(L/2)\>$ as a function
of the fraction of the chain occupied by the block, $\ell/L$ using the same
data. All the points collapse to a single scaling function, which we fit to a
CFT finite-size form \cite{Calabrese.JSTAT.04}
\begin{equation}
  \label{S.fs.cft}
  \tilde S(\ell) = S(\ell) - S(L/2) \approx {c \log(2) \over 3} \log\[ \sin\(
  \pi {\ell\over L} \) \],
\end{equation}
and plot the resulting curve along with the points. The difference between the
fitting curve and the points is apparent, so we proceed to substract them and
plot the result in the inset of figure (\ref{fig.Sgs}, bottom). The residual
appears to correspond to {\em higher harmonics}, showing that a different
scaling function, $Y(x)$, is required to account for the finite-size effects
\cite{Fagotti.PRB.11}. The Fourier series representation of that function can
be written as
\begin{equation}
  \label{fagotti.fourier.expansion}
  Y(x) = \[ 1 + \sum_{j=1}^{\infty} k_j\] \sin{x} - \sum_{j}^{\infty}
  \frac{k_j}{2j+1} \sin{[(2j+1)x]}
\end{equation}
and the more general expression for the finite-size average von Neumann
entropy is given by
\begin{equation}
  \label{fagotti.correction}
  S(\ell) \approx {c\log{(2)}\over 3} \log{\[ {L\over \pi} Y\(
    \pi{\ell\over L} \) \]} + c'
\end{equation}
the contribution of the first modes provides a good approximation to the
entropy. Fitting the finite-size data to this new functional form, we find the
additive constant $c'\approx 0.7338$ and the amplitude of the first mode
$k_1=0.1025$, which are close to the value $c'\approx 0.726$ reported by
Laflorencie and \cite{Laflorencie.PRB.05} and the value $k_1=0.115$ obtained
by Fagotti et al. \cite{Fagotti.PRB.11}.

Despite the many similarities between the average behavior of entanglement in
the random hopping model and a conformally invariant system in 1D, there are
also substantial differences. One of the most relevant is in the R\'enyi
entropies. In the conformal case they present characteristic parity
oscillations \cite{Xavier.PRB.11,Calabrese.JPA.09}:
\begin{eqnarray}\nonumber
  S_\alpha(\ell) &\approx& \frac{c}{6} \(1+\frac{1}{\alpha} \)
  \log{\[\frac{L}{\pi} \sin\( \pi \frac{\ell}{L} \) \]} +
  c^{\prime}\\ \label{renyi.osc} 
  &+& (-1)^{\ell} f_\alpha \[\frac{L}{\pi} \sin{\(\pi \frac{\ell}{L}\)}\]^
  {-2K/ \alpha}
\end{eqnarray}
where $c$ and $c'$ are the same as in eq. (\ref{S.fs.cft}), $f_\alpha$ is the
oscillation amplitude, which typically increases with $\alpha$, $K$ is the
Luttinger parameter ($K=1$ in our case) and the term $(-1)^{\ell}$ corresponds
to $\cos(2k_F \ell)$, where $k_F=\pi/2$ is the Fermi moment for
half-filling. On the other hand, within the bond-structure picture, all
R\'enyi entropies are equal to the von Neumann case since, in the strong
disorder regime, an $\ell$-size block has $2^\ell$-fold degenerate eigenvalues
$2^{-\ell}$ \cite{Refael_Moore.JPA.09} then, the $\alpha$ order R\'enyi
entropy is

\begin{equation}
\label{eq:RenyEQvN}
S_\alpha = \frac{1}{1-\alpha} \log{ 2^{\ell (1-\alpha)}} = \log {2^\ell}
\end{equation}

\begin{figure}[h!]
  \centering
  \resizebox{8.0cm}{!}{\includegraphics{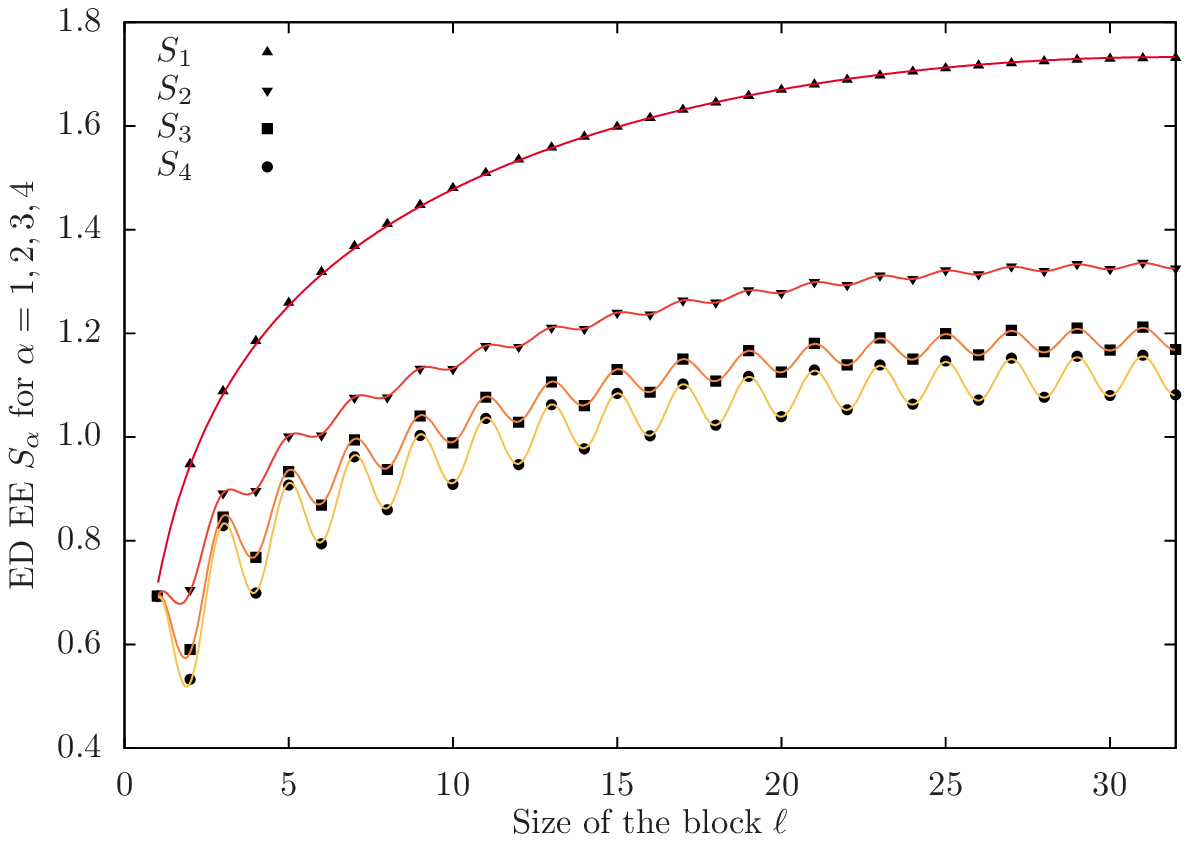}}
  \resizebox{8.0cm}{!}{\includegraphics{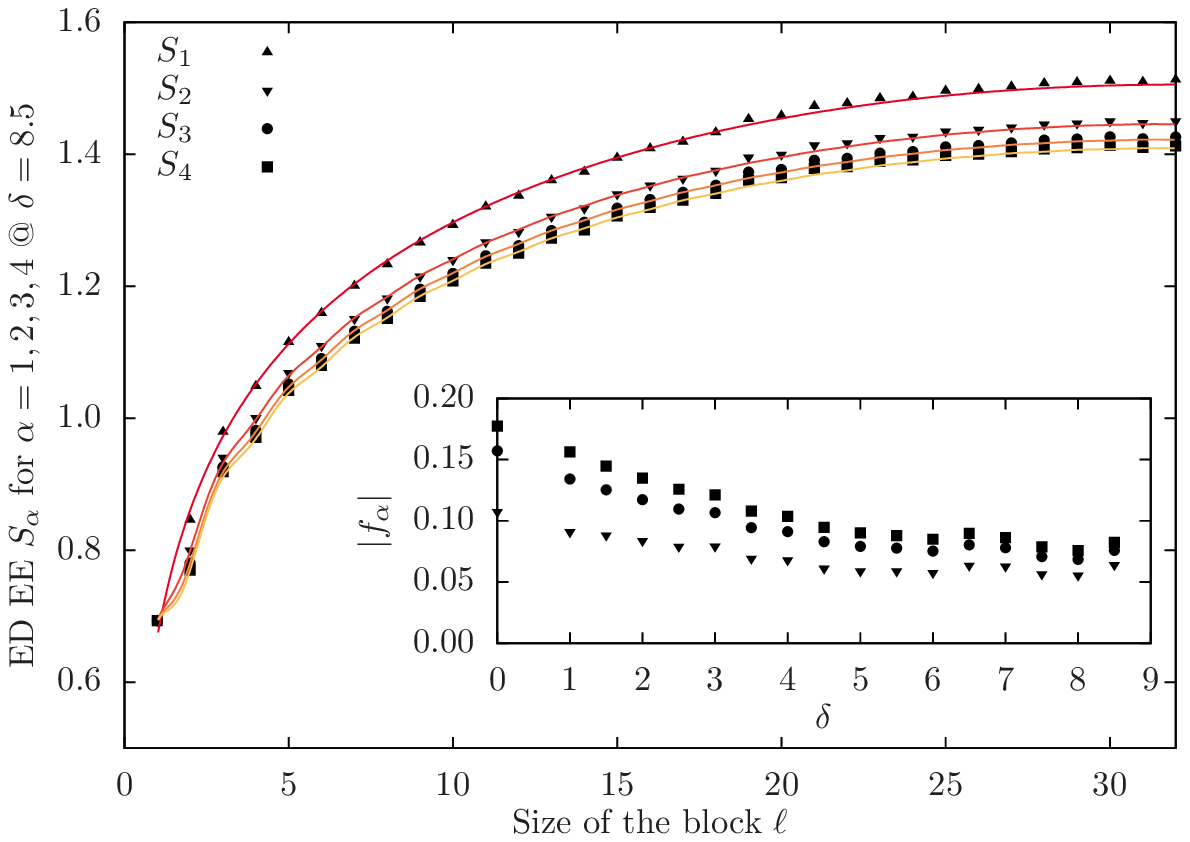}}
  \caption{Top: R\'enyi block entropies in the clean case, for $L=64$ and
    PBC. Notice the strong parity oscillations in the higher order
    entropies. Bottom: average R\'enyi entropies for $2\cdot 10^4$
    realizations of a $L=64$ system with $\delta=8.5$. Notice how entropies of
    all orders become much closer, and how the oscillations attenuate. The
    inset shows a decrease in the magnitude of the oscillation amplitude
    $f_\alpha$ in eq. (\ref{renyi.osc}) as a function of $\delta$.}
  \label{fig.renyi}
\end{figure}

Figure (\ref{fig.renyi}) compares the average R\'enyi entropies obtained with
exact diagonalization in the clean and strongly disordered cases for a system
with $L=64$, $\delta=8.5$ using $2\cdot 10^4$ disorder realizations, for the
lowest R\'enyi orders ($\alpha$ from $1$ to $4$). The upper panel shows the
clean case, notice the strong parity oscillations in the higher order R\'enyi
entropies. The bottom panel of depicts the average R\'enyi entropies in the
disordered case. Notice that their amplitude is substantially lower. The inset
in the lower panel of figure (\ref{fig.renyi}) analyses that decrease in
amplitude: the $f_\alpha$ factors fitted in eq. (\ref{renyi.osc}) are plotted
against $\delta$, the disorder intensity. They can be seen to attenuate very
slowly. In fact, even for very large $\delta$ they are still not negligible
showing that the conjecture in \cite{Calabrese.etal.PRL.10}
holds. Nonetheless, for infinite disorder, the effect of $f_\alpha$ will
disappear.

\subsection{Variance of the von Neumann entropy}

\begin{figure}[t]
  \centering
  \resizebox{8.0cm}{!}{\includegraphics{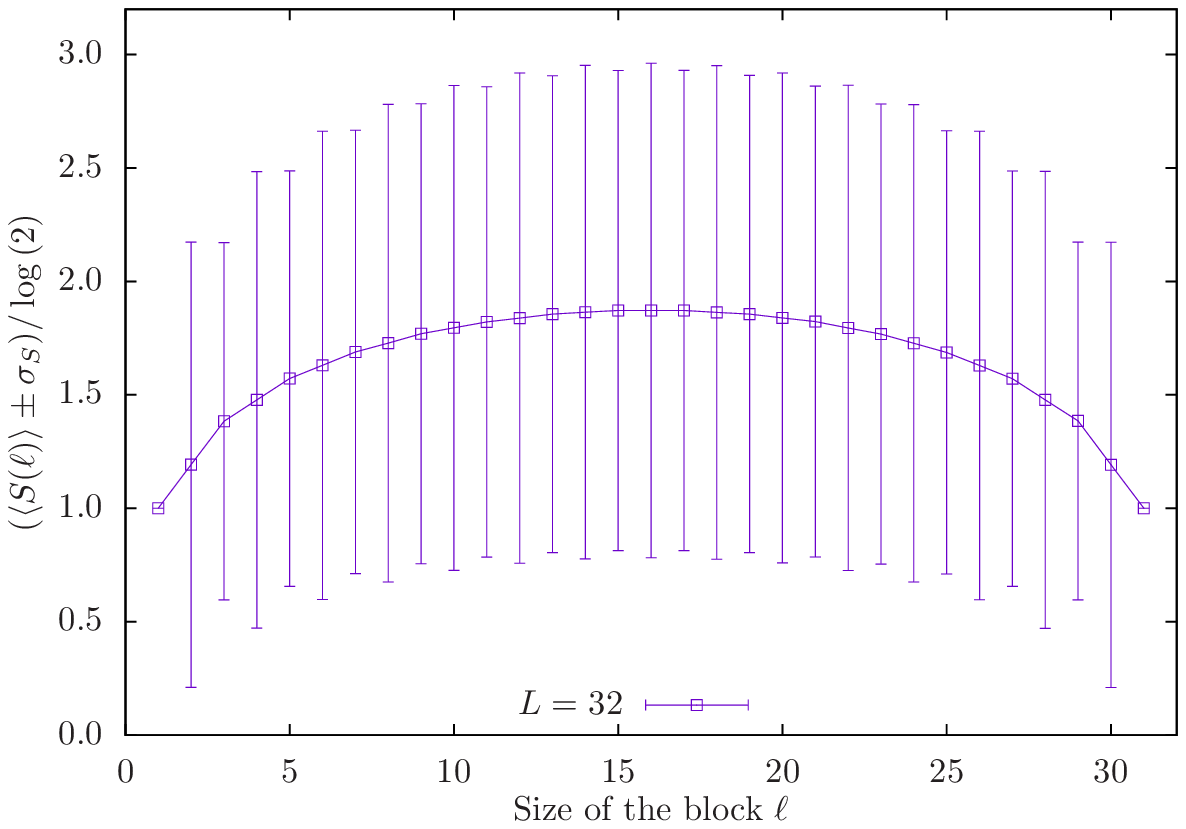}}
  \resizebox{8.0cm}{!}{\includegraphics{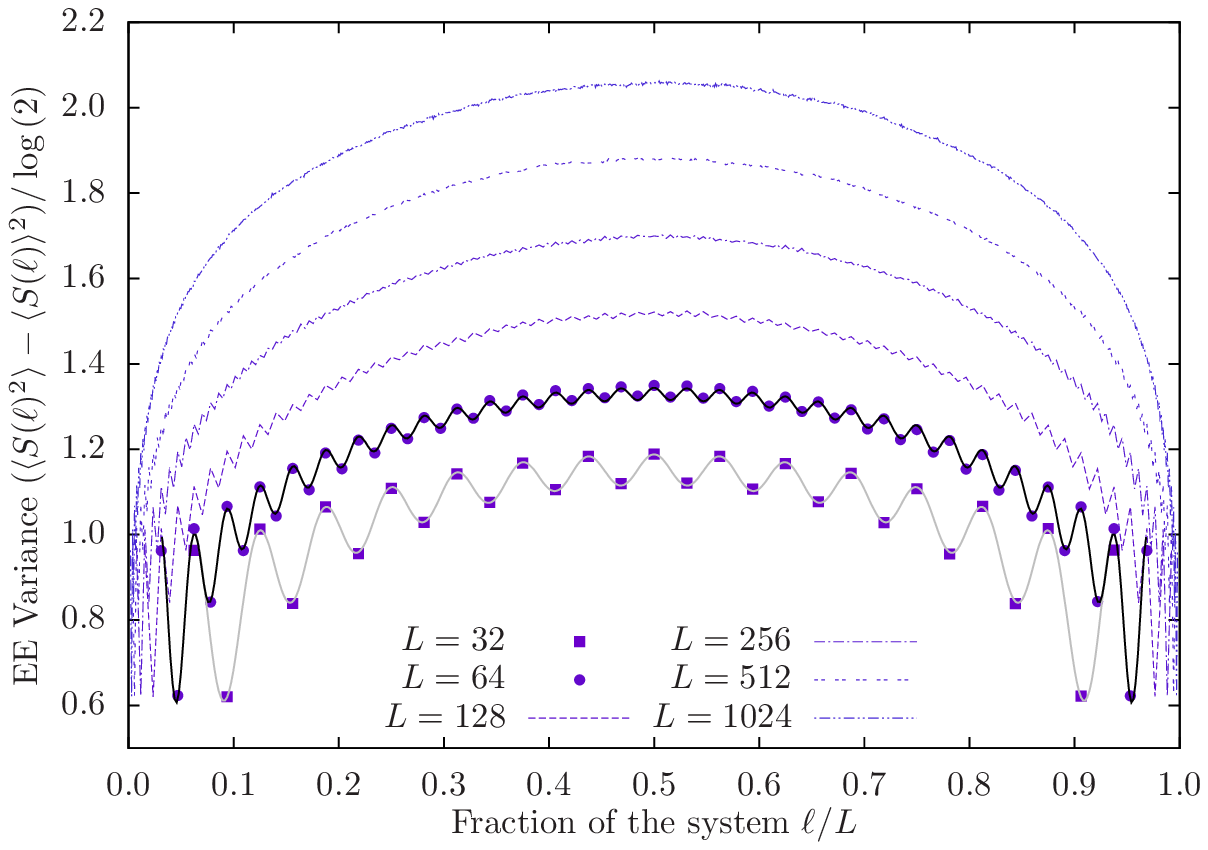}}
  \caption{Top: average von Neumann entropy for $\delta=8$ and $L=32$,
    obtained with the RG, with errorbars given by the standard
    deviation. Notice the parity oscillation in the errorbars, which are
    larger for even blocks. Bottom: variance of the von Neumann entropy
    distribution for different sizes and $\delta=8$, obtained with the
    RG. Notice how the parity oscillations fit accurately expression
    (\ref{fit.variance}).}
  \label{fig.variance}
\end{figure}

The variance of the von Neumann entropy presents also interesting universal
behavior alike to the CFT predictions, but with an interesting difference:
parity oscillations remain even in the strong disorder regime, as the RG
calculations show. Fig. (\ref{fig.variance}) depicts the results of
simulations run with $10^6$ samples for sizes $L=32$, $64$, $128$, $256$,
$512$ and $1024$, obtained with the RG and $\delta=8$, along with a very
accurate fit to a law similar to (\ref{renyi.osc}):
\begin{eqnarray}\nonumber
  \sigma^2_{S} &=& c_\sigma \log(2) \log{\[\frac{L}{\pi} \sin\( \pi
    \frac{\ell}{L} \) \]} + c'_\sigma \\ \label{fit.variance}
  &+& (-1)^{\ell} f_\sigma \[\frac{L}{\pi} \sin{\(\pi
    \frac{\ell}{L}\)}\]^{-2K_\sigma} 
\end{eqnarray}
with $c_\sigma\approx 0.4$, $c'_\sigma\approx 0.46$, $f_\sigma\approx 0.78$
and $K_\sigma\approx 2/3$. Remarkably, the oscillations are also present in
the higher order cummulants of the distribution. They are only absent in the
average.

The origin of those oscillations in the variance of the von Neumann entropy,
and their accurate fit to the CFT expression is an open problem. These
oscillations bear resemblance to the density oscillations found by
\cite{Song.PRB.10} in a clean system, which are explained as an effect of the
boundaries and subleading corrections to the CFT prediction. 

Notice that the variance is always higher for the even blocks, and the
even-odd difference is much larger for smaller blocks.  Also let us remark
that although the average number of outgoing bonds increases smoothly as we
increase the block size, the probability distributions are quite different:
even-sized blocks can only cut an even-number of bonds, and viceversa.

\subsection{Open boundary conditions}

\begin{figure}
  \centering
  \resizebox{8.0cm}{!}{\includegraphics[angle=270]{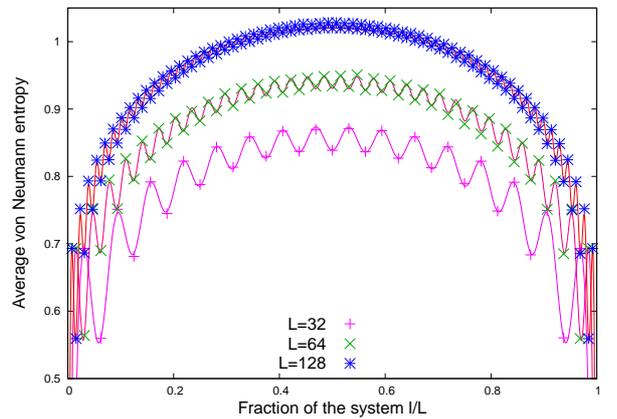}}
  \caption{Average von Neumann entropy of the random hopping model with open
    boundary conditions. Notice the characteristic parity oscillations, which
    fit to a Luttinger parameter $K=1/2$. }
  \label{fig.open}
\end{figure}

Let us consider what are the differences in the case of open boundary
conditions. In that case, translational invariance is lost: the entropy of a
block depends not only on its size, but also on its distance to the extreme of
the chain. It is customary to choose blocks starting from the left extreme. In
that case block only presents one inner boundary instead of two. The CFT
prediction for the clean (critical) case is that the prefactor of the
logarithmic term in the expression of the von Neumann entropy is halved. In
the disordered case we can also observe a reduction of the entanglement
entropy, but with remarkable differences. Figure (\ref{fig.open}) shows the
average von Neumann entropy for three sizes ($L=32$, $64$ and $128$) with open
boundary conditions, using $10^6$ realizations with $\delta=8$. Notice the
parity oscillations, which are similar to those appearing in the higher order
R\'enyi entropies with periodic boundary conditions. In fact, a fit to a
expression similar to (\ref{renyi.osc}) works very well \cite{Taddia.PRB.13}
\begin{eqnarray}\nonumber
  S(\ell) &\approx& {c_{open} \log(2) \over 6} \log{\[ {L\over \pi} \sin\( \pi
    {\ell\over L} \) \]} + c'_{open} \\\label{s.open} 
  &+& (-1)^\ell f_{open} \[\frac{L}{\pi} \sin{\(\pi \frac{\ell}{L}\)}\]^ {-K_{open}}
\end{eqnarray}
where $c_{open}\approx 1.5$, $c'_{open} \approx 0.76$, $f_{open}\approx -0.24$
and $K_{open}\approx 1$. Thus, even though the entropy is reduced in the case
of open boundary conditions, the results in this case differ considerably from
the expectation that $c_{open}$ should be one, but the value for $K_{open}$
agrees with previous results obtained for clean systems \cite{Dalmonte.PRB.11,
  Calabrese.etal.PRL.10}.

\subsection{Odd chains}

\begin{figure}
  \centering
  \resizebox{8.0cm}{!}{\includegraphics[angle=270]{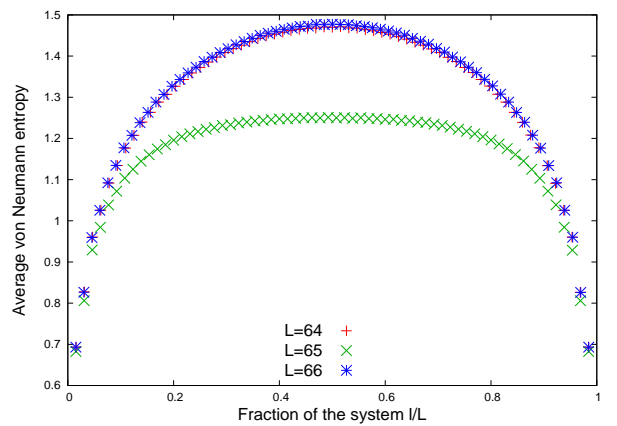}}
  \caption{Average von Neumann entropy compared for even and odd number of
    sites. Notice the {\em plateau} which gets established for intermediate
    block sizes in the case of odd chains.}
  \label{fig.odd}
\end{figure}

On the other hand, disordered chains present very different behavior when the
number of sites is {\em odd}, as opposed to the clean case. Effectively, in
that case one site is not allowed to establish a bond, and entanglement is
effectively reduced, see fig. (\ref{fig.odd}). Moreover, bonds can not be
established over the single site and, thus, this site can be regarded as an
{\em opening} in the boundary conditions. Effectively, the average von Neumann
entropy becomes nearly flat for intermediate block sizes, showing a {\em
  plateau}.

\section{The Bond-Length Distribution}
\label{sec:bondpic}

Let us consider a 1D random hopping chain of length $L$ and PBC, close enough
to the IRFP, where the bond-structure picture becomes accurate to describe the
ground state of the system. Given a bond between sites $i_1$ and $i_2$, let
$l_b\equiv|i_1-i_2|$ (mod $L$) be its length. Let us consider the probability
distribution for the bond lengths, $P(l_b)$. As indicated in
\cite{Hoyos.PRB.07}, we will show that all entanglement properties in the IRFP
stem from the knowledge of this $P(l_b)$ and the assumption of (approximate)
bond independence, beyond the constraint that two bonds can never cut.

The scaling behavior of $P(l_b)$ has been estimated via the Dasgupta-Ma RG
\cite{Fisher.PRB.94}. As the RG proceeds, the typical length scale of the
bonds increases. It can be argued that the likelihood of a given site
surviving until the typical length scale is $l_b$ scales as $l_b^{-1}$. A bond
can be established only between two surviving sites, so, if we assume
independence, the probability of establishing a bond of length $l_b$ scales as
the product: $P(l_b)\approx l_b^{-2}$. If that probability distribution is
assumed to be exact for all (odd) values of $l_b$, the normalization constant
should be $8/\pi^2$. But for small values of $l_b$ the fitting exponent
deviates from $-2$. For the scaling regime, the best fit is found to be
$P(l_b) \approx (2/3)\; l_b^{-2}$ \cite{Hoyos.PRB.07}.

\begin{figure}[h!]
  \resizebox{8.0cm}{!}{\includegraphics{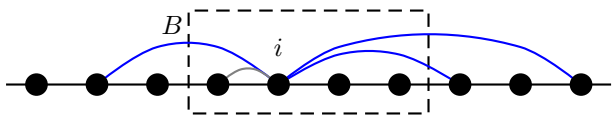}}
  \caption{Illustration of the bond counting procedure which leads to
    expression (\ref{hoyos.1}). Let us consider box $B$ of size $\ell$. Bonds
    stemming from site number $i$ will contribute to the block entanglement
    only if their length is larger than their distance to the boundary (blue
    bonds only). Notice that actual bonds are only allowed if their length is
    {\em odd}.}
  \label{fig.box}
\end{figure}

The average von Neumann entropy of a block $B$ with size $\ell$ is given by
the expected number of bonds crossing its boundaries, multiplied by $\log(2)$,
see fig. (\ref{fig.box}). Let the sites in the block be numbered from $1$ to
$\ell$, and consider site $i$ and its bond. Let $i'$ be the other extreme. The
bond will contribute to the entropy if its length is larger than the distance
to the boundary. If $i'$ is at the left of $i$, then the bond only contributes
if $l_b \geq i$. The expected number of such bonds is $\sum_{l_b=i}^{L/2}
P(l_b)$. If $i'$ is at the right of $i$, the bond will contribute if $l_b\geq
\ell-i+1$, and we get $\sum_{l_b=\ell-i+1}^{L/2} P(l_b)$. Summing for all $i$,
and considering that leftwards and rightwards bonds are equally likely, we
get:
\begin{equation}
  \label{hoyos.1}
  S_1(\ell) = {\log(2) \over 2} \sum_{i=1}^\ell \[ \sum_{l_b=i}^{L/2} P(l_b) +
  \sum_{l_b=\ell-i+1}^{L/2} P(l_b) \].
\end{equation}

This expression can be recollected into a more convenient one
\cite{Hoyos.PRB.07}:
\begin{equation}
  \label{hoyos.formula}
  S(\ell) = \log{(2)} \[ \sum_{l_b=1}^\ell l_b P(l_b) + \ell
  \sum_{l_b=\ell+1}^{L/2} P(l_b) \]
\end{equation}
where the first term is the most relevant, since smaller bonds have the
largest probabilities. Inserting the previous estimate for $P(l_b)\approx
(2/3)\;l_b^{-2}$ into the first term of eq. (\ref{hoyos.formula}), we obtain
$S(\ell) \approx (\log(2)/3)\; \log(\ell)$, as in eq. (\ref{S.disorder}).

\begin{figure}[h!]
  \centering
  \resizebox{8.0cm}{!}{\includegraphics{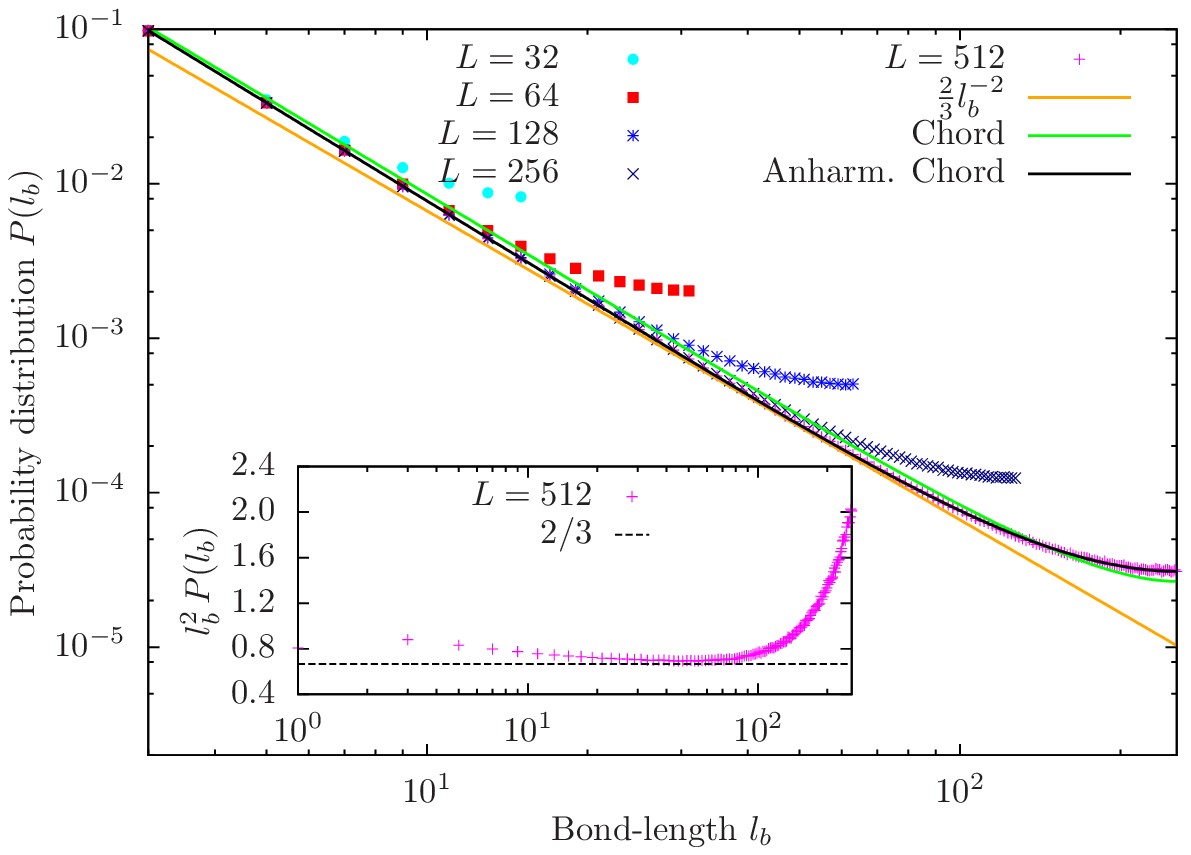}}  
  \resizebox{8.0cm}{!}{\includegraphics{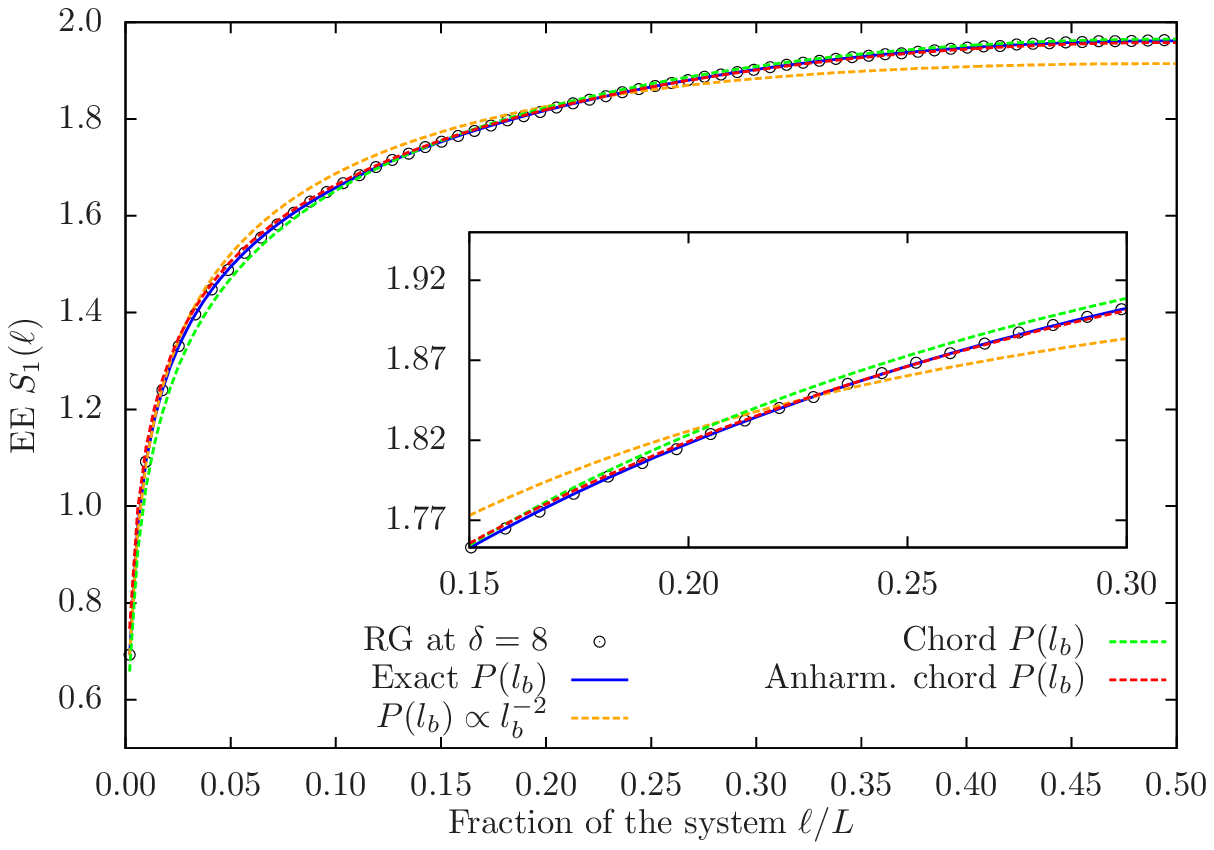}}
  \caption{Top: Probability distribution for the bond-lengths obtained with
    the RG and $\delta=8$ for different sizes. Alongside, the scaling fit to
    intermediate bond-lengths, $(2/3)\;l_b^2$, and the fits to the chord
    --eq. (\ref{chord.law})-- and the anharmonic chord
    --eq. (\ref{anharmonic.chord}). Inset: plot of $l_b^2 P(l_b)$, showing the
    approach to $2/3$ during the scaling regime. Bottom: average von Neumann
    entropy for $L=512$ and $\delta=8$ (dots), along with predictions obtained
    by inserting different approximations to $P(l_b)$ in
    eq. (\ref{hoyos.formula}): the scaling law $P(l_b)\propto l_b^{-2}$, the
    chord approximations, and the exact $P(l_b)$ obtained from the
    simulations. Notice that the accurate fit for this last one, validating
    expression (\ref{hoyos.formula}). Inset: detail of the same plot.}
  \label{fig.pls}
\end{figure}

Figure (\ref{fig.pls}, top) studies the behavior of $P(l_b)$ by averaging over
five million disorder realizations with $L$ ranging from $32$ to $512$ and
$\delta=10$, in logarithmic scale. The leading $l_b^{-2}$ behavior is
apparent, as a fit for intermediate values of $l_b$ shows. The straight line
corresponds to the scaling regime approximation, $P(l_b)=(2/3)\;
l_b^{-2}$. The large-$l_b$ deviation, for $l_b$ comparable to the system size,
is a finite-size correction. Let us distribute the $L$ points uniformly in a
circumference of diameter $1$. The probability for a bond between sites
separated $l_b$ lattice units is approximately proportional to the inverse
squared of their actual distance, i.e., to their {\em chord}:
\begin{equation}
  \label{chord.law}
  P(l_b) \propto \sin^{-2} \(\pi l_b/L \).
\end{equation}

The accuracy of the fit to $P(l_b)$ can be further improved using an {\em
  anharmonic} chord approximation:
\begin{equation}
  \label{anharmonic.chord}
  P(l_b)\propto \[ Y(l_b) \]^{-\gamma}
\end{equation}
with $Y(l_b)$ given in expression (\ref{fagotti.fourier.expansion}), and only
retaining the first anharmonic term. The fit gives $k_1=0.12$ and
$\gamma=2.11$, with very good accuracy.

Figure (\ref{fig.pls}, bottom) compares the average entropy $S(\ell)$ obtained
by direct sampling with three possible estimates from the probability
distribution for the bond-lengths using eq. (\ref{hoyos.formula}): (i) the
scaling law $P(l_b)\propto l_b^{-2}$, (ii) the chord and anharmonic-chord
laws, eqs. (\ref{chord.law}) and (\ref{anharmonic.chord}), and (iii) the
sampled distribution for $P(l_b)$. Notice that approximation (iii) is
indistinguishable from the sampled entropy.

It is interesting to ask whether the bond-length samples are actually
independent or not. We have investigated the bond-length correlations. Given a
bond-structure, consider the list of the bond-lengths obtained when the bonds
are ordered according to the index of their left-most site: $\{l_{b,1},
l_{b,2}, \cdots, l_{b,L/2}\}$. Let us consider the conditional probabilities
$P(l_{b,i}|l_{b,i-1})$, i.e.: the probability of finding a bond-length
$l_{b,i}$ knowing that the previous bond-length was $l_{b,i-1}$. The
independence assumption is equivalent to $P(l_{b,i}|l_{b,i-1}) =P(l_b)$, i.e.:
that knowledge of the previous bond-length is irrelevant. In fact, this
assumption is {\em false}. After a bond of length $l_{b,i-1}=3$, a bond
$l_{b,i}=1$ must ensue. Nonetheless, the difference $|P(l_{b,i}|l_{b,i-1})
-P(l_{b,i})|$ decays to zero very fast when $l_{b,i-1}$ grows. Since the
contribution to the entropy is larger for larger bonds, the independence
assumption becomes accurate in that case.

\subsection{Order Statistics for the Bond-Length}

\begin{figure}[t]
  \centering
  \resizebox{8.0cm}{!}{\includegraphics{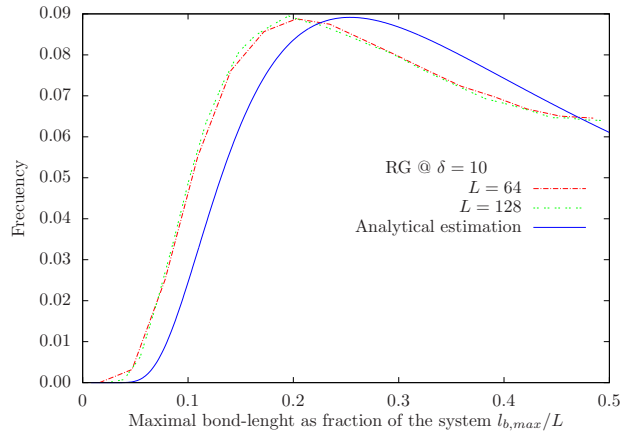}}
  \caption{Histogram for $l_{b,max}/L$ in the IRFP. Notice how both curves
    seem to converge to a thermodynamic limit. For low $l_{b,max}$, the
    probability increases fast up to a a value $l^M_{b,max}/L$, which is close
    to $0.2$. The continuous curve corresponds to the estimate for
    $\pi_1(l_b)$ given in equation (\ref{pi1.estimate}).}
  \label{fig.lmax}
\end{figure}

Let $\pi_k(l_b)$ denote the probability distribution function (PDF) for the
$k$-th longest bond. Thus, $\pi_1(l_b)$ will be the PDF for the longest bond
in the system, $l_{b,max}$. Figure (\ref{fig.lmax}) shows the histogram found
over five million realizations with $\delta=10$ for $L=32$ and $L=64$ with the
RG. A thermodynamic limit curve appears for those relatively small sizes, with
a peak at $l_{b,max}/L \approx 0.2$, i.e.: the longest bond covers
approximately $1/5$ of the total system. After the maximal bond, the curve
appears almost flat, up to $1/2$, which is the maximal realizable value.

The independence assumption allows us to give an estimate for
$\pi_1(l_b)$. Let $X$ be a 1D random variable with probability distribution
$P(X)$, and $X_{max,N}$ represent the maximal observation out of a series of
$N$ independent realizations. The probability distribution for $X_{max,N}$ can
be found this way: (i) find the cumulative distribution function (CDF) for
$X$: $F(x) \equiv P(X>x) = \int_{-\infty}^x p(s)\;ds$; (ii) the CDF for the
maximal observation is just $P(X_{max,N}>x)=F(x)^N$; (iii) the probability
distribution for the maximal observation is found by differentiation of the
CDF: $P(X_{max,N})=\partial_x \[ F(x)^N \]$. Since we have $L/2$ bonds in our
system, the CDF for the maximal bond will be $F(l_b)^{L/2}$. Assuming a
continuous PDF $P(l_b)\propto l_b^{-2}$, we get the estimate
\begin{equation}
  \label{pi1.estimate}
  \pi_1(l_b) \propto \(1-{1\over l_b}\)^{L/2-1} \( {1\over l_b^2} \).
\end{equation}

This estimate, which is plotted in fig. (\ref{fig.lmax}), can be used to find
the value of $l^M_{b,max}$, the most likely maximal bond-length. In the
thermodynamic limit, $l^M_{b,max} \approx L/4$.

\begin{figure}
  \centering
  \resizebox{6.0cm}{!}{\includegraphics[angle=270]{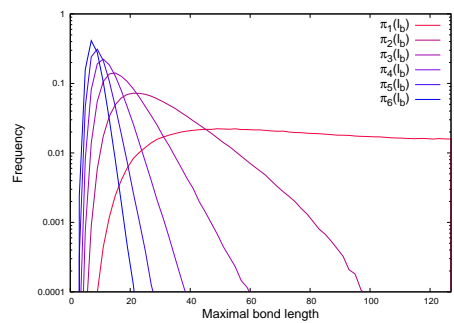}}
  \caption{Probability distribution functions for the $k$-th longest
    bond in a chain with $256$ sites, with $\delta=10$ and $10^6$
    realizations.}
  \label{fig.longbonds}
\end{figure}

Figure (\ref{fig.longbonds}) depicts the different $\pi_k(l_b)$, i.e.: the PDF
for the $k$-th longest bond, for a system with $256$ sites and $10^6$ disorder
realizations. Notice how they become more and more peaked as $k$ increases.

\subsection{Longest bond and energy gap}

The average energy gap $\Delta E$ is known to vanish very fast in the
thermodynamical limit \cite{Fisher.PRB.94}. In this section we will consider
the relation between this average gap and entanglement. Since energy scales
are linked to length scales, the connection is made via the longest bond. We
have applied the Dasgupta-Ma RG to $10^6$ disorder realizations with
$\delta=8$ for systems of $L=64$, $128$ and $256$. The average value of
$\log(\Delta E)$ for each value of $l_{b,max}$ fits to an exponential decay:
\begin{equation}
  \label{loggap}
  \<\log(\Delta E)\> \approx A + B \exp(-l_{b,max}/l_0)
\end{equation}
In all three cases, $l_0\approx L/5$, i.e.: the expected value for the maximal
bond-length.

\begin{figure}
  \centering
  \resizebox{6.0cm}{!}{\includegraphics[angle=270]{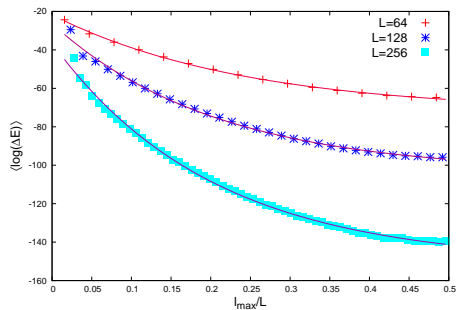}}
  \caption{Average logarithm of the energy gap as a function of the
    $l_{b,max}$ in each realization, for three system sizes ($L=64$, $128$ and
    $256$), and $\delta=8$. Alongside, fit to functional form (\ref{loggap}).}
  \label{fig.gaps}
\end{figure}

\section{Random permutations and entanglement}
\label{sec:perm}

A simple model can be devised which reproduces most features of the ground
state of the random hopping model, in which all the disorder effects are
collected into a model of {\em random permutations}.

Let us consider a variant of the random hopping model in which each new
disorder realization is associated with a random permutation $\sigma$ of the
set $\{1,\cdots,L\}$. Let us associate the $i$-th element of the permutation,
$\sigma_i$, to the $i$-th hopping term of the chain: $J_i
=\exp(-\sigma_i)$. The rationale is that the renormalization rule
eq. (\ref{renormalized.coupling}) becomes now additive in the values of
$\sigma_i$:
\begin{equation}
  \label{rg.permutations}
  \sigma^{(R)}_i = \sigma_{i+1}+\sigma_{i-1}-\sigma_i,
\end{equation}
i.e.: the lowest element of the permutation is removed, along with its two
neighbors, and all three are replaced by a renormalized element. Each random
permutation determines a bond-structure, which in turn determines all the
correlation and entanglement properties within the ground state of the
system. Thus, we conjecture that sampling over disorder realizations amounts
to sampling over random permutations, i.e.: a discrete set of
possibilities. Random permutation theory has already made appearance in other
areas of physics, such as the statistical mechanics of growing interfaces
\cite{Krug.JPA.10}, where it links the shape fluctuations in the
Kardar-Parisi-Zhang (KPZ) universality class with the Tracy-Widom probability
distributions from random matrix theory.

\begin{figure}
  \centering
  \resizebox{8.0cm}{!}{\includegraphics{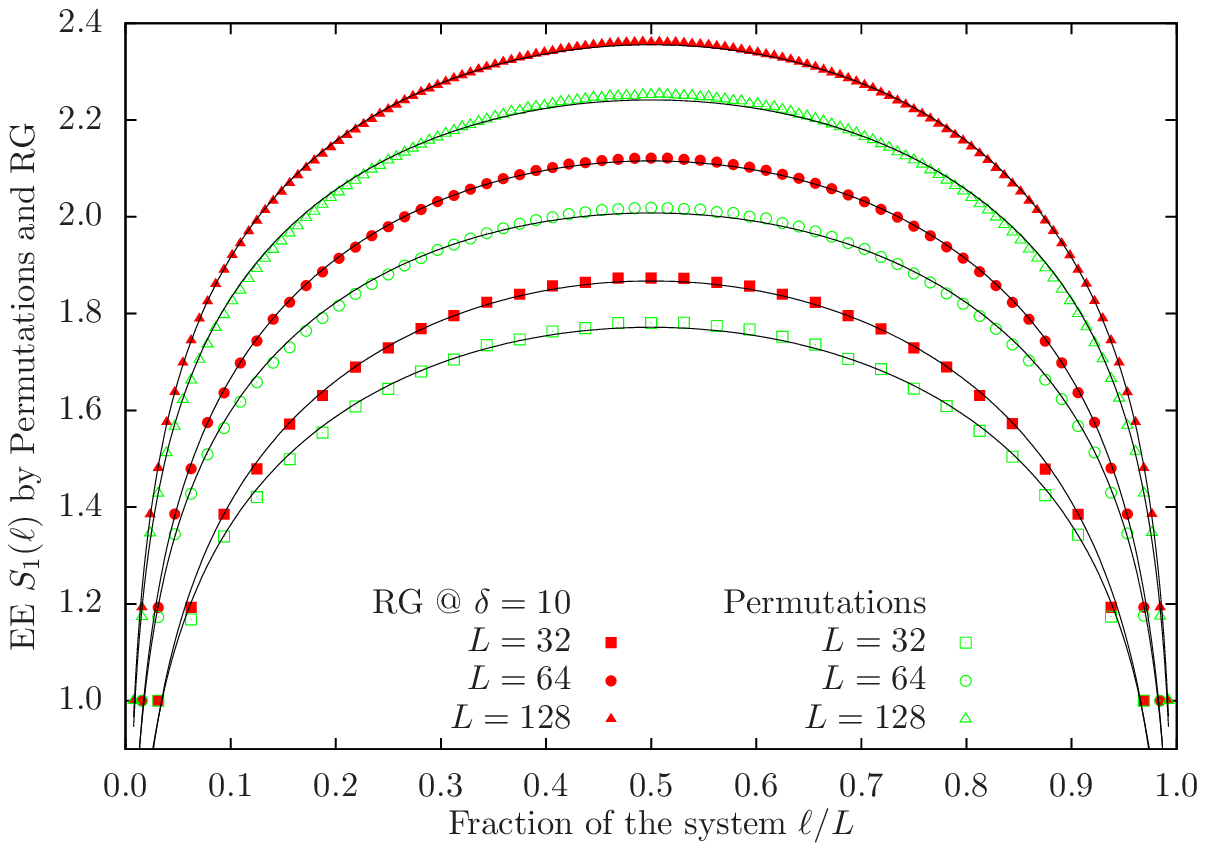}}
  \resizebox{8.0cm}{!}{\includegraphics{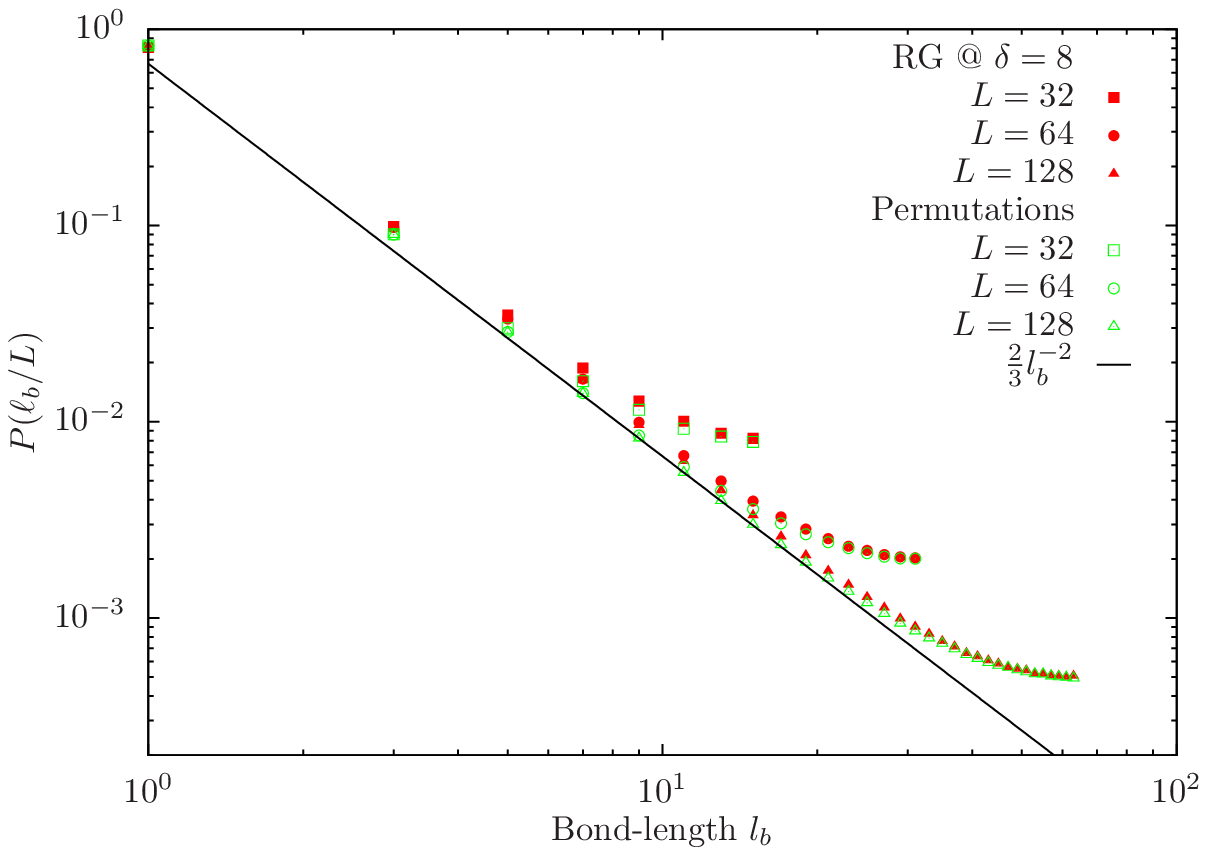}}
  \resizebox{8.0cm}{!}{\includegraphics{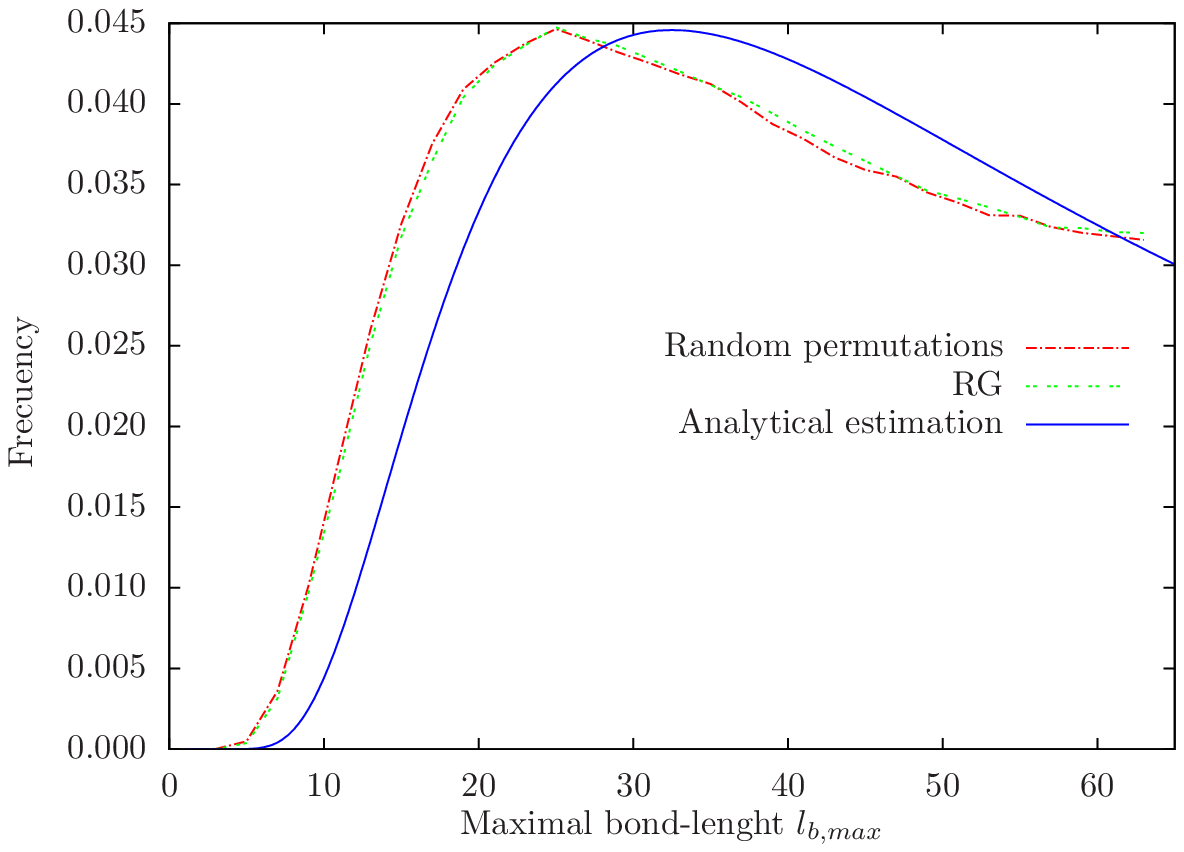}}
  \caption{Study of entanglement of the ground state of the random
    permutations model. Top: average von Neumann entropy as a function of
    $\ell/L$, for $L=32$, $64$ and $128$. Center: Histogram for the bond
    lengths, $l_b$. Bottom: Histogram for $l_{b,max}/L$ obtained by RG and
    random permutations for $L=128$. Notice how both curves seem to converge
    to a thermodynamic limit. For low $l_{b,max}$, the probability increases
    fast up to a a value $l^M_{b,max}/L$, which is close to $0.2$.}
  \label{fig.lmax.perm}
\end{figure}

Our RG flow in the permutation space is not perfectly determined. It sometimes
finds {\em coincidences}, i.e.: despite all elements are initially different,
after some RG steps, some of them will coincide. If the coinciding elements
are sufficiently far apart, the order in which we renormalize them is
immaterial. In a few cases, they are close enough, thus forcing to choose one
of them randomly in order to proceed. Nonetheless, those coincidences get more
and more sparse as the system size grows, and become negligible in the
thermodynamic limit.

Let us show that the Dasgupta-Ma RG and the random permutations model give the
same results for the entanglement. As it was discussed above, all the relevant
magnitudes stem from a single function: the probability distribution for
$P(l_b)$. Figure (\ref{fig.lmax.perm}) shows runs performed for $10^5$ samples
for $L=32$, $64$ and $128$ for the average von Neumann entropy (top), bond
length histogram (center) and maximal bond length histogram (bottom), along
with comparison with the Dasgupta-Ma RG approach.

\begin{table}[h!]
  \centering
  \begin{tabular}{|c|c|c|c|c|c|c|}
    \hline
    &\multicolumn{3}{c}{Permutations}&\multicolumn{3}{|c|}{RG}\\
    \cline{2-7}
    $L$  &$c$    &$c'$   &$\chi^2/10^{-4}$&$c$    &$c'$   &$\chi^2/10^{-4}$\\
    \hline 
    $32$ &$1.043$&$0.567$&$2.2$          &$1.167$&$0.519$&$2.3$\\
    $64$ &$1.047$&$0.557$&$1.4$          &$1.148$&$0.524$&$1.0$\\
    $128$&$1.048$&$0.547$&$1.2$          &$1.124$&$0.538$&$0.5$\\
    \hline
  \end{tabular}
  \caption{Fitting  values for  the von  Neumann  entanglement entropy
    (see eq (\ref{S.fs.cft}))  to  compare  the   model  of  random
    permutations and  the RG method for $\delta=10$  and $5\cdot 10^6$
    samples.}
  \label{tab:permutations}
\end{table}

The main feature of the random permutations model is the strong hierarchy
among the link strengths. Our model bears strong similarities to the {\em
  hierarchical model} of RNA-folding \cite{Muller.PhD,Wiese.JSTAT.08}. In this
model, random binding energies are provided for each pair of sites on a 1D
chain, and bonds are established among them in order with a no-crossing
condition. Renormalization group arguments show that the universality class is
captured merely by choosing the $L(L-1)/2$ binding energies $\epsilon_{ij}$
such that $\epsilon_{i_1j_1} \ll \epsilon_{i_2j_2} \ll \cdots \ll
\epsilon_{i_{L(L-1)/2} j_{L(L-1)/2}}$. This model can be considered an
infinite-dimensional version of the random hopping model.

\section{Excited states}
\label{sec:exc}

Entanglement of excited states has been studied recently within the CFT
framework \cite{Alcaraz.PRL.10, Ibanez.JSTAT.12, Taddia.PRB.13,
  Dalmonte.PRB.12, Eloy.PRB.12, Essler.PRL.13}. In this section we will extend
the techniques developed in our work to study the excited states of the random
hopping model. Indeed, entanglement of all eigenstates can be constructed
using either exact diagonalization, Dasgupta-Ma RG or random permutations
approaches. As it has been stated above, the eigenstates of the hopping matrix
constitute the single-body modes: bonds between pairs of sites, with negative
energy, and their corresponding anti-bonds, with positive energy. The ground
state is obtained by filling up the set of all negative energy modes, i.e.:
all the bonds. The full spectrum of the Hamiltonian is obtained as we either
reduce the number of particles to allow empty modes and/or add particles in
modes with positive energy. Both negative and positive energy modes, bonds and
anti-bonds, give the same contribution to the entanglement entropy but, when
both are present on the same pair of sites, their contribution to entanglement
cancels out, leaving two factorized sites.

For a clean system, the entanglement entropy increases substantially when a
particle-hole (PH) excitation is created, i.e.: when a particle in an occupied
mode is upgraded to an empty state above the Fermi level
\cite{Ibanez.JSTAT.12,Essler.PRL.13}. Moreover, entanglement remains invariant
for {\em compact states}, i.e.: states in which the list of occupied modes
presents no holes. Those states are represented by vertex operators.

\begin{figure}
  \centering
  \resizebox{3.8cm}{!}{\includegraphics{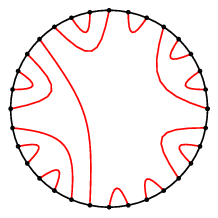}}\hspace{0.2cm}%
  \resizebox{3.8cm}{!}{\includegraphics{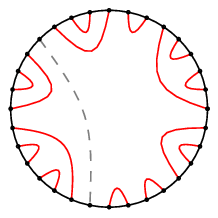}}\vspace{0.2cm}
  \resizebox{3.8cm}{!}{\includegraphics{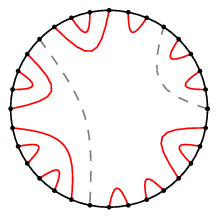}}\hspace{0.2cm}%
  \resizebox{3.8cm}{!}{\includegraphics{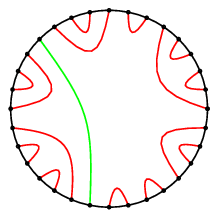}}
  \caption{Pictorial representation of the excited states. Top-left:
    Bond-structure of the ground state $\ket|0>$. Top-right: the excited state
    $\ket|1>$ is obtained by removing the longest bond. Bottom-left: if the
    second longest bond is removed, the excited state $\ket|2>$ is
    obtained. Bottom-right: the PH state is built by upgrading the closest
    particle to the Fermi point to the first mode above it. Due to the
    particle-hole symmetry, in our case we upgrade the longest bond to the
    corresponding anti-bond, which presents the same entanglement.}
  \label{fig.EXpict}
\end{figure}

The situation is very different for the strongly disordered system. Figure
(\ref{fig.EXpict}) illustrates the different types of excited states and their
effects on entanglement. In the top-left panel we show a possible bond
structure describing the ground state. The lowest energy excitation is the
compact state obtained by either removing the weakest bond or adding a
particle on the weakest anti-bond. Both cases result in the longest bond being
removed from the system, as shown in the top-right panel. A second compact
excitation can be obtained by removing/adding a further particle, as shown in
the bottom-left panel. The last panel shows the effect of a PH excitation, in
which the longest bond is upgraded to be an anti-bond, which leaves the
entanglement structure untouched.

\begin{figure}
  \centering
  \resizebox{8.0cm}{!}{\includegraphics{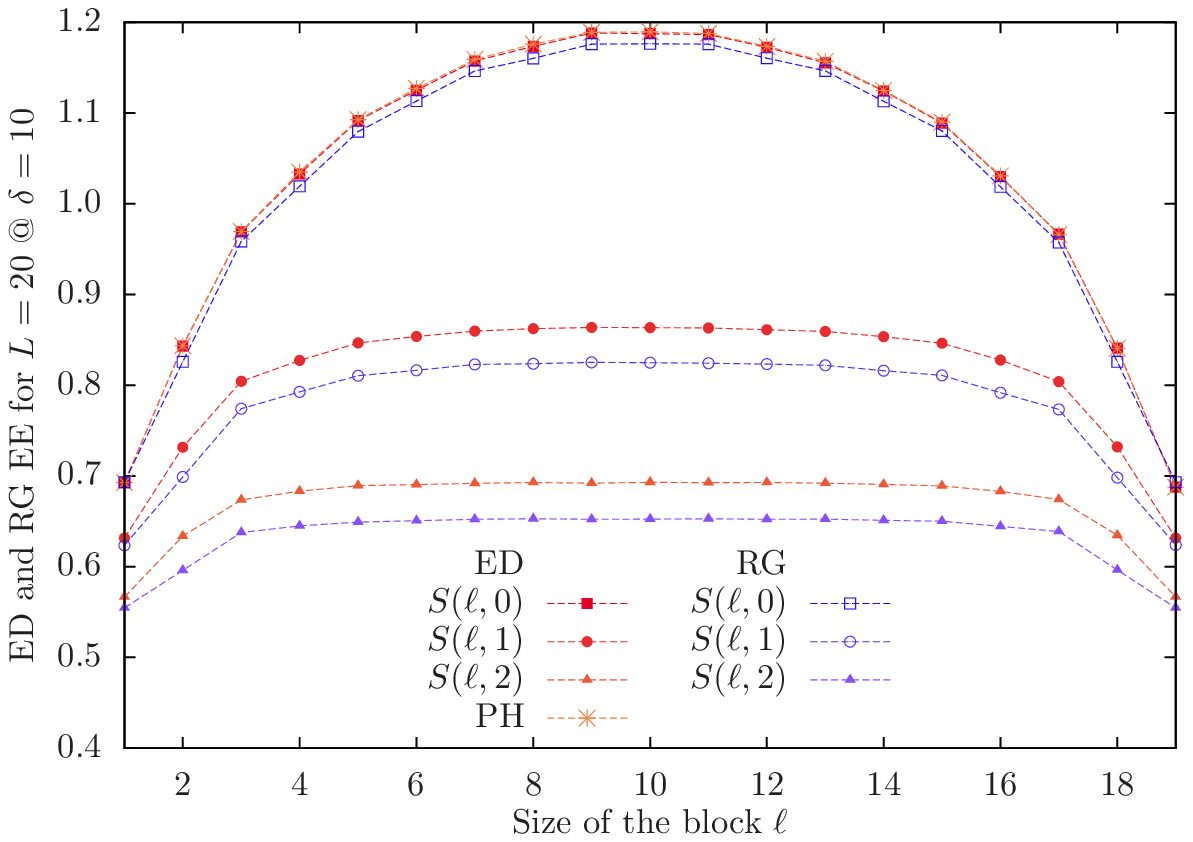}}
  \resizebox{8.0cm}{!}{\includegraphics{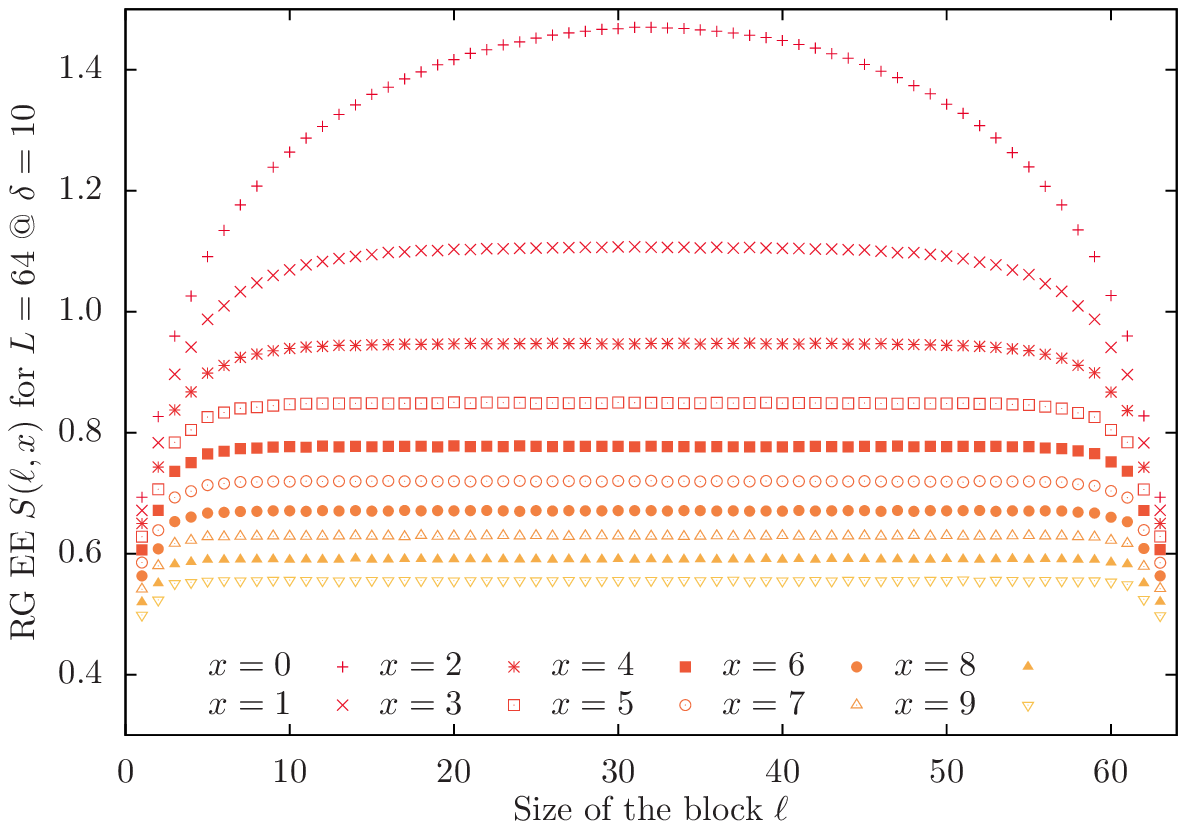}}
  \caption{Top: Average von Neumann entropy of the ground state and low-energy
    excited states, both with exact diagonalization and RG. Note the decrease
    of the entropy in the case of excited states $\ket|1>$ and $\ket|2>$, and
    the invariance for the PH excitation. Bottom: $S(\ell,x)$ for a system of
    $64$ sites for different number of removed/added particles $x$ ($x=0$
    corresponds to the ground state).}
  \label{fig.excited.ed}
\end{figure}

Let $\ket|x>$ denote the excited state in which $x$ particles have been
removed from the ground state (equivalently, we could say added), and let
$S(\ell,x)$ denote the average von Neumann entropy of a block of size $\ell$
within state $\ket|x>$. Figure (\ref{fig.excited.ed}, top) shows this average
von Neumann entropy for the ground state and three excited states, obtained
with exact diagonalization and the RG. The first one, the PH excitation,
coincides with the entanglement of the ground state. The other two correspond
to states $\ket|1>$ and $\ket|2>$, in which one ($S(\ell,1)$) or two
($S(\ell,2)$) particles are added or removed. Notice that, in this case, a
{\em plateau} appears for intermediate block sizes, similar to the one
appearing for the ground state of odd-sized systems. The bottom panel of
fig. (\ref{fig.excited.ed}) shows how this plateau reduces slowly its height
as the number of added/removed particles increases, i.e.: the curves
$S(\ell,x)$ flatten progressively for increasing $x$.

\begin{figure}[h]
  \centering
  \resizebox{8.0cm}{!}{\includegraphics{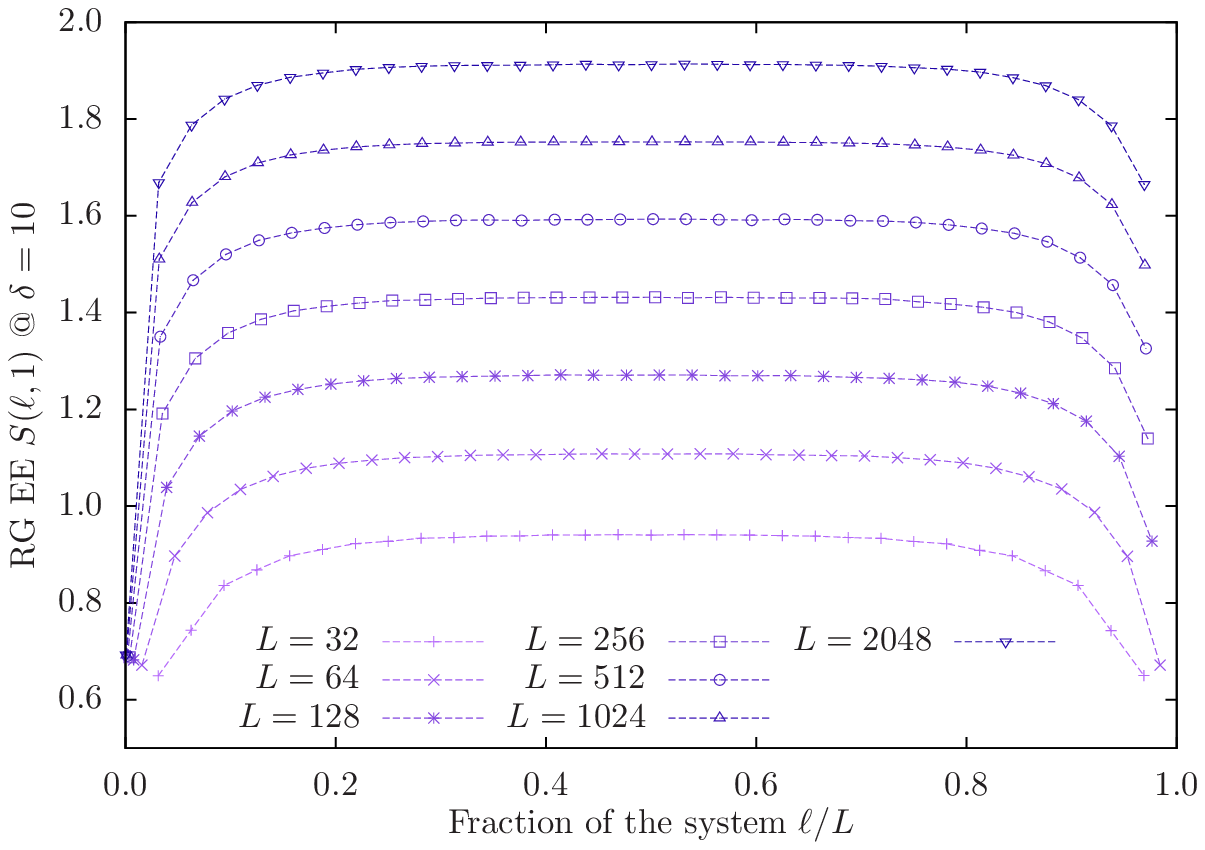}}
  \resizebox{8.0cm}{!}{\includegraphics{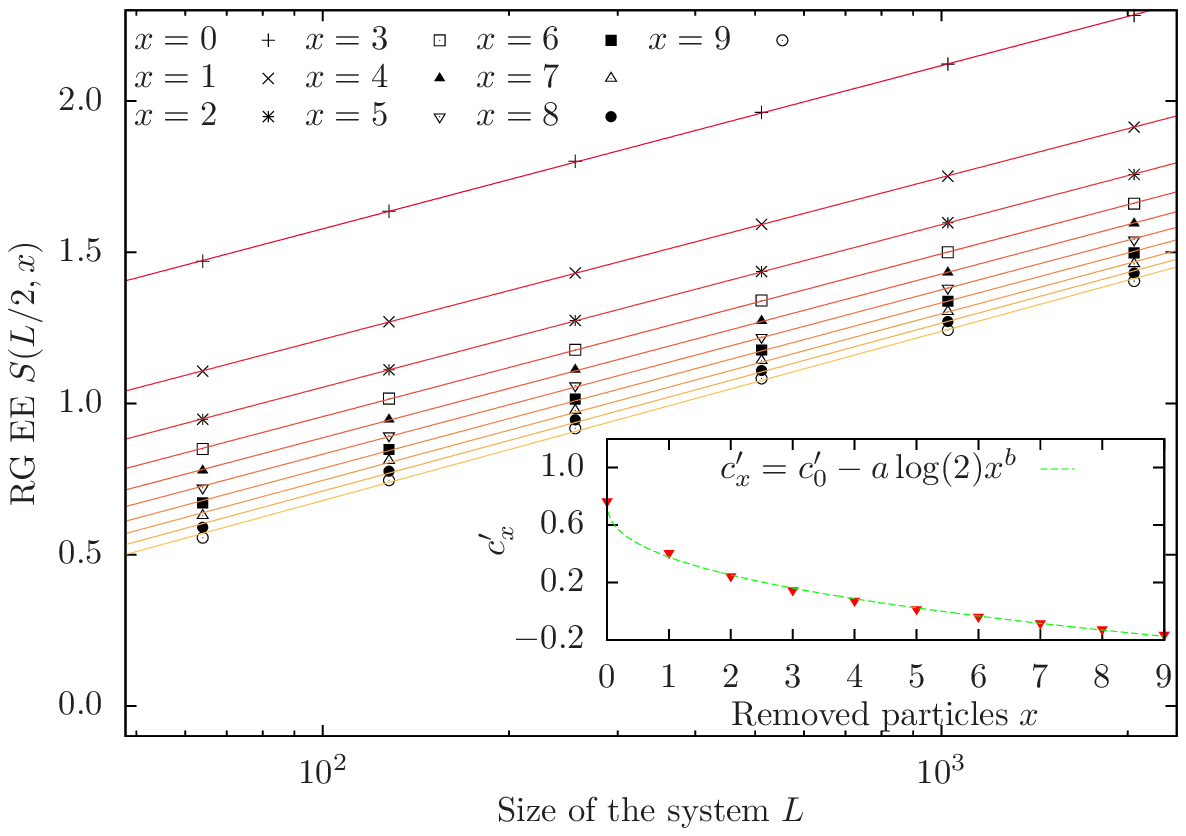}}
  \caption{Top: $S(\ell,1)$ for different system sizes $L=32$, $64$, $128$,
    $256$, $512$ and $1024$. Notice that while the sizes grow geometrically,
    the maximal values of the entropy grow only arithmetically. Bottom: height
    of the {\em plateau} of $S(L/2,x)$, as a function of the system size, in
    solid lines represent the fit to expression (\ref{S.fs.cft}), and the
    inset shows the additive constant $c_{ex}'(x)$ as function of the removed
    particles $(a=0.56,b=0.4)$.}
  \label{fig.excited}
\end{figure}

Figure (\ref{fig.excited}, top) shows the behavior of $S(\ell,1)$ for sizes
ranging from $L=32$ to $L=2048$, as obtained with the RG. All of them present
a similar plateau, but at increasing heights. Notice that the sizes are in
geometric progression, and the plateau heights appear to grow only
arithmetically. This shows that the behavior of $S(L/2,1)$ is be {\em
  logarithmic} with the system size $L$. Indeed, let us claim that

\begin{equation}
S(L/2,x)={c_{ex} \log(2) \over 3} \log(L) + c'_{ex}(x).
\label{sl2}
\end{equation}
with $c_{ex}=1$. This claim receives support from the results shown in the
bottom panel of fig. (\ref{fig.excited}), which shows $S(L/2,x)$ as a function
of $L$ (in logarithmic scale), for different values of $x$. Notice that all
curves are, in fact, parallel straight lines, and the slope is indeed close to
$\log(2)/3$. The additive constant $c'_{ex}(x)$ is the only difference, and
its decay with $x$ is shown in the inset of fig. (\ref{fig.excited}). For the
$x=1$ case, the reduction in the value of the additive constant from the
ground state can be explained by assuming a reduced {\em effective system
  size}, from $L$ to $L/5$, i.e.: $c'_{ex}(1) \approx c' -
\log(2)\log(5)/3$. This reduction in the effective system size can be
explained if we assume that it coincides with the {\em length of the expected
  maximal bond}.

The curves $S(\ell,1)$ for different sizes collapse when the maximum value is
substracted from the entropy values and, in the thermodynamic limit, they fit
to the finite-size form:
\begin{equation}
  \label{Sex.finitesize}
  S(\ell,1) \approx \frac{c\log{(2)}}{3}\log(L) +c_x'
  -\beta e^{-\gamma\sqrt{\ell/L}}
\end{equation}
where $\gamma \approx 10$ allows us to estimate the size of the region in
which entropy grows to reach the value $S(L/2,1)$ as approximately $1/5$ of
the total size of the system. Remarkably, $L/5$ is again the average size of
the bond of maximal length, as shown in figure (\ref{fig.lmax}).

From all those analysis we can attempt a physical picture of the entanglement
in the first excitation. Removal of the weakest bond is usually the same as a
removal of the longest bond, which has a typical size $l_{b,max}\approx
L/5$. Since bonds can not cross, entanglement can grows normally only {\em
  within} the region of size $\approx L/5$ which lies under this longest
bond. The region outside, with size $4L/5$, is devoid of long bonds, and
contributes less to the entanglement. Similar arguments apply for the higher
excitations.

\section{Conclusions}
\label{sec:conclusions}

In this work we have analyzed the properties of entanglement in random hopping
models, focusing on the similarities between the CFT predictions for the clean
case and the strong disorder RG predictions. We have used a combination of
exact diagonalization, the Dasgupta-Ma renormalization scheme and a new tool
based on the study of random permutations. All techniques coincide in
providing a compelling image, based on a bond-picture.

All the entanglement properties within the ground state stem from the
probability distribution for the bond lengths $P(l_b)$ and an assumption of
approximate independence for large bond lengths. Both the thermodynamic limit
and the finite-size form for the average von Neumann entropy can be deduced
from the scaling analysis of that distribution function. The behavior of the
R\'enyi entropies can not be established solely from the bond picture, since
the Dasgupta-Ma RG prediction is that all orders will give the same
result. Indeed, we have observed that the parity oscillations which appear in
the clean case, according to the CFT prediction, attenuate as the disorder
grows, making them similar for all values of the R\'enyi order.

Interesting results were obtained for odd chains, where a {\em plateau}
appears in the average von Neumann entropy, for intermediate system
sizes. Moreover, parity oscillations appear both in the average von Neumann
entropy of chains with open boundary conditions and in the {\em variance} of
the von Neumann entropy in all cases. They fit nicely an expression similar to
the CFT prediction, but with different constants. Remarkably, the scaling of
the maximal variance is again logarithmic, but with a different prefactor.

We have introduced the {\em random permutation picture}, which is a
simplification of the Dasgupta-Ma renormalization in which the hoppings are
given fixed values which differ broadly in order of magnitude, but are
distributed at random among the lattice links. All the properties of
entanglement and correlation in the ground state can be established solely in
this picture.

Furthermore, we have analysed the average entanglement of excited
states. Indeed, excited states are of two types: those which convert a
negative energy mode into its corresponding positive energy mode do not alter
the bond picture. But excitations which add or remove particles alter them in
a remarkable way. Indeed, the average entanglement entropy of the first
excitation presents a {\em plateau} at intermediate sizes, whose magnitude
scales logarithmically with the size of the system {\em as if} the size of the
system corresponds to the average size of the maximal bond. Higher excitations
result in a further reduction of the effective size of the system.

There are still many open questions related to this system. First of all, a
thorough analysis of entanglement for intermediate values of the disorder would
clarify the decay and convergence of the parity oscillations in the R\'enyi
entropies of all orders. Moreover, it would be interesting to study the system
from a dynamical point of view, i.e.: the clean to disorder transition.

\begin{acknowledgments}
  We would like to acknowledge F.~C. Alcaraz and J.~A. Hoyos for very useful
  discussions, along with CNPq for the hospitality at The S\~ao Carlos
  Institute of Physics (Brazil). We also acknowledge financial support from
  the Spanish government from grant FIS2012-33642. J.R.-L. acknowledges
  support from grant FIS2012-38866-C05-1.
\end{acknowledgments}


\end{document}